\documentclass{aastex61}               
\usepackage{amsmath,amssymb,amstext,wasysym}
\usepackage{gensymb}
\usepackage{ulem}
\usepackage{longtable}

\usepackage[all]{hypcap} 

\usepackage{natbib}
\usepackage{multirow}
\shorttitle{Forming Mercury by Giant Impacts}
\shortauthors{Chau et al.}

\begin{document}

\title{Forming Mercury by Giant Impacts}
\author{Alice Chau$^1$,  Christian Reinhardt$^1$, Ravit Helled$^1$, Joachim Stadel}
\affil{Center for Theoretical Astrophysics and Cosmology, Institute for Computational Science, University of Zurich,
Winterthurerstrasse 190, CH-8057 Z\"urich, Switzerland}

\begin{abstract}

The origin of Mercury's high iron-to-rock ratio is still unknown. 
In this work we investigate Mercury's formation via giant impacts and consider the possibilities of a single giant impact, a hit-and-run, and multiple collisions in one theoretical framework. 
We study the standard collision parameters (impact velocity, mass ratio, impact parameter), along with the impactor's composition and the cooling of the target. It is found that the impactor's composition affects the iron distribution within the planet and the final mass of the target by up to 15\%, although the resulting mean iron fraction is similar. 
We suggest that an efficient giant impact requires to be head-on with high velocities, while in the hit-and-run case the impact can occur closer to the most probable collision angle (45$^{\circ}$). 
It is also shown that Mercury's current iron-to-rock ratio can be a result of multiple-collisions, with their exact number depending on the collision parameters. 
Mass loss is found to be more significant when the collisions are tight in time. 
\end{abstract}

\section{Introduction}
Mercury has a unique composition. Its mean density is similar to the Earth's but Mercury is 20 times lighter and cannot be subject to the same self-compression. 
This suggests the existence of a large metallic core consisting $\sim$ 70\% of the planet's mass, i.e., a large iron-to-rock ratio (hereafter $Z_{Fe}$) which is about twice the proto-solar abundance (e.g., \citealt{Spohn}, \citealt{Hauck2013}). 
Several scenarios have been proposed to explain Mercury's high $Z_{Fe}$ and they cover different stages of the planet formation process and rely on different chemical and physical mechanisms. 
One class of mechanisms is linked to the separation of metals and silicates in the solar nebula. This can be a result of different condensation temperatures for metals and silicates  \citep{Lewis}, their different conductive properties and reaction to photophoretic forces \citep{Wurm13}, or a result of different gravitation and drag force balance \citep{Weidenschilling}. 
These mechanisms however, typically require specific disk architecture and conditions. 

A second class of scenarios suggests that Mercury lost a large fraction of its rocky mantle after its formation by evaporation followed by solar wind \citep{Cameron} or mantle stripping by a giant impact \citep{Benz1988}. 
Gravitational collisions between bodies of similar sizes are very common in the final stages of planet formation, (e.g., \citealt{Chambers},\citealt{Quintana}). 
Collisions can include violent giant impacts that are energetic enough to strip away part of the mantle, and one giant impact is sufficient to reach Mercury's current $Z_{Fe}$ \citep{Benz2007}. 
However, the exact conditions that lead to  Mercury's formation via a giant impact are still unknown. Recent scenarios proposed that Mercury might have collided with another body as large as Earth or Venus \citep{Asphaug}. 
Most simulations of the late formation stage start with embryos of Mars size as it is presently still computationally challenging to resolve the inner disk in such simulations.  
To our knowledge only \citet{Lykawka} have investigated the formation of Mercury in an inner disk simulation. In the best terrestrial planet systems, the analogs tend to be slightly heavier than Mercury's current mass leaving open the possibility of a giant impact formation scenario for Mercury.

The giant impact hypothesis has been under debate since the MESSENGER mission results (\citealt{Peplowski}, \citealt{Ebel2017}) where relatively high abundance of potassium (K), thorium (Th) and uranium (U) were measured. 
The initial interpretation of these measurements suggested that any scenario involving high temperatures, or high energies, which would remove the volatiles from Mercury is excluded.
This argument is based on our knowledge of the Moon's composition. The Moon, which is also thought to form  via a giant impact, is volatile depleted with a K/Th ratio $\sim$5 times lower than the Earth's.  
However, the volatile depletion of the Moon is linked to recondensation and not vaporization (\citealt{Stewart2016}, \citealt{Lock2018}). 
Unlike Mercury, the Moon probably formed from a debris disk that accumulated to form the satellite. 
By analogy to the Earth-Moon system, Mercury is the Earth remnant, which retains a significant amount of volatiles. 
In addition, the disrupted silicate material in Mercury could be well-mixed and preserve its original composition \citep{Nittler2017}. 
Fractionation of volatiles within the condensed silicates is not expected, but some volatiles could be lost from atmospheres/oceans \citep{Schlichting2015, Genda2005}. There may also be transfer between impactors in hit-and-run events \citep{Burger2018}. Generally, giant impacts might not lead to volatile depletion for terrestrial planets, as argued by \cite{Ebel2017}, despite their very different impact histories, they seem to have very similar K/Th and K/U ratios.

In this paper we investigate the giant impact hypothesis. We consider (1) a single giant impact (2) a hit-and-run and (3) multiple collisions in one numerical framework. We investigate a large parameter space for individual collisions to understand the outcome possibilities, as well as the sensitivity to the impactor's composition and proto-Mercury's initial state. We find that all three options can lead to the formation of a Mercury-like planet, although each scenario requires different impact conditions. 

This paper is organized as follows. In Section \ref{methods} we describe the methods used to model the planetary bodies and their respective collisions. We also explain the tools for the analysis of the simulation outcomes such as the clump finder. In Section \ref{Paramterstudy} we present the results of the collisions of the parameter space we have explored. In Section \ref{resultsmulti} we discuss our approach to the multiple collision scenario. In section \ref{discussion} we briefly discuss the importance of following the evolution of the ejecta and impactor, and compare our results with previous studies. In Section \ref{conclusion} we summarize and discuss the results. 

\section{Methods \label{methods}}
\subsection{Smooth-Particles-Hydrodynamics Methods}
To model the two-body collisions we use the smoothed particle hydrodynamics (SPH) code \textsc{Gasoline} \citep{Wadsley}, a modern SPH implementation that was adapted for planetary collisions \citep{Reinhardt}.
The colliding bodies are assumed to be composed of condensed materials that are modeled with the Tillotson equation of state (EOS) \citep{Tillotson}.  
While the Tillotson EOS lacks a thermodynamically-consistent treatment of mixed phases and phase transitions\footnote{The Tillotson EOS does not follow the phase transitions. For example, for the liquid-vapor phase it linearly interpolates the pressure between a low-density liquid and the gaseous phase.}, it has a simple analytical form that can easily be implemented in SPH simulations (e.g., \citealt{Benz1987}, \citealt{Canup2001}, \citealt{Genda2012}, \citealt{Marinova2011} or \citealt{Burger2018}). In addition, the  Tillotson EOS shows a good agreement with measured data \citep{Brundage2013} 
as well as with more thermodynamically consistent EOS such as ANEOS \citep{Thomson1972}. 
The good agreement, however, is limited to the relatively low-velocity collisions where only a small fraction of the material is close to the vaporization heat of the rocks (\citealt{Emsenhuber2018}, \citealt{Canup2004}, \citealt{Benz1989}), where the proper treatment of phase transitions can affect the thermal pressure. 
While these EOS differences are important for the Moon-forming collisions where one is interested in the detailed physical states of the orbiting material, in case of a Mercury-stripping impact, where we concentrate on the total mass, these differences are expected to have a small influence on the results. 
The bodies are assumed to be fully differentiated with a chondritic abundance, i.e., an iron core (30\%) and a basalt mantle (70\%). The particle representation of the planets are generated as described in \citet{Reinhardt}, in order to allow for multi-component bodies the procedure was slightly modified as described below. First a 1D equilibrium model is obtained by solving the structure equations with boundary conditions $ M(r=R_{CMB})=M_{core}$, $M(r=R_p)=M_p$ and $\rho(r=R_p)=\rho_0$ where $R_{CMB}$ and $R_p$ are the radius of the core-mantle boundary and the planet, respectively. At the CMB we assume temperature and pressure to be continuous. 
The thermal profile is adiabatic. For an initial guess of the density and internal energy in the core, $\rho_c$ and $u_c$ are varied until the above boundary conditions are satisfied. Then the SPH particles are arranged on concentric shells, where each shell is divided using an equal area tessellation of the sphere in order to obtain a very uniform distribution. For each material the particle distribution is generated separately in order to properly capture the transition between them. The resulting initial conditions (IC) are closely following the model and show very low noise, thus the particles are very close to the equilibrium configuration. Because standard SPH can not properly capture the density discontinuity at the core-mantle boundary (e.g., \citealt{Canup2001}) we still evolve the planets in isolation for a few hours in simulation time until the root mean square velocity is below 50m/s. {We show an example of proto-Mercury in Fig \ref{DensityProfile}.}
We use an intermediate resolution (80k to 270k particles in total) in order to explore a large parameter space. The resolution of proto-Mercury is kept constant (55k particles) while the impactor's resolution is adapted such that all SPH particles have the same mass. All simulations lasted for 2.2 days in simulation time until the fragment mass converged.
SPH results depend on the numerical resolution (e.g. \citealt{Hosono2016}, \citealt{Reinhardt}). In our simulations, the masses of the post-impact bodies converge with high resolution $N=10^6$ within $\sim$ 5-10\%. 

\subsection{Parameter Space}
A giant impact is characterized by the following parameters: the impact parameter $b$, the relative velocity between the two bodies $v_{imp}$, and the mass of the target and the impactor. 
We focus on the regime in the parameter space that leads to mantle stripping. 
Our baseline models begin with a proto-Mercury with a mass of $\geq$ 2.25~M$_{\mercury}\equiv$~M$_{mpM}$\footnote{Note that here we provide the initial conditions in units of the proto-Mercury's minimal mass, while in the rest of the paper we discuss using Mercury's mass.}, which is the minimal mass assuming chondritic abundance needed to obtain Mercury's current core mass. 
Since we find that some of the iron from the core can be lost we also consider cases with higher masses, of 1.1, 1.2~M$_{mpM}$
We consider various masses for the colliding body: 0.1, 0.3, 0.4, 0.5, 2, 3, 5~M$_{mpM}$.  The relative velocities are in the range of  $v=[10, 60]$~km/s, typically a few times the escape velocity of the system. For a pairwise collision, the  escape velocity is given by
\begin{equation}
v_{esc}=\sqrt{\frac{G(M+m)}{R+r}},
\end{equation}
where $M$, $R$, and $m$, $r$ are the mass and radius of the first and second body, respectively.
For reference, Mercury's escape velocity is $\sim$4 km/s. 
The impact parameter $b$ is taken to be between 0.1 and 0.7. A very small $b$ is not considered since in that 
case either the energy is too low to eject material from the target and the impactor is eventually accreted, or the energy is too high and both bodies are destroyed. 
A large $b>0.7$ is also irrelevant since such grazing impacts do not efficiently strip the mantle. 

We denote as ``Case-1" the collisions where proto-Mercury is the target which is hit by a smaller body (such as in \citealt{Benz2007}). ``Case-2" refers to the hit-and-run case where proto-Mercury is actually the impactor and collides with a larger body (such as in \citealt{Asphaug}) which no longer resides in the Solar System. 
To determine Mercury's mass after the impact, we use the clump finder SKID \citep{Stadel} that uses a closest-neighbor-algorithm and temperature and pressure comparison to determine if a given SPH particle is part of a clump bound by gravity. For an individual clump, this is equivalent to testing whether the surrounding SPH particle's velocities are smaller than the escape velocity. Depending on the simulation, Mercury's mass is either the largest (Case-1) or second largest remnant (Case-2).

\section{Results: Giant Impact Simulations \label{Paramterstudy}}

\begin{figure}[h!]
\begin{center}
\includegraphics[width=0.24\textwidth]{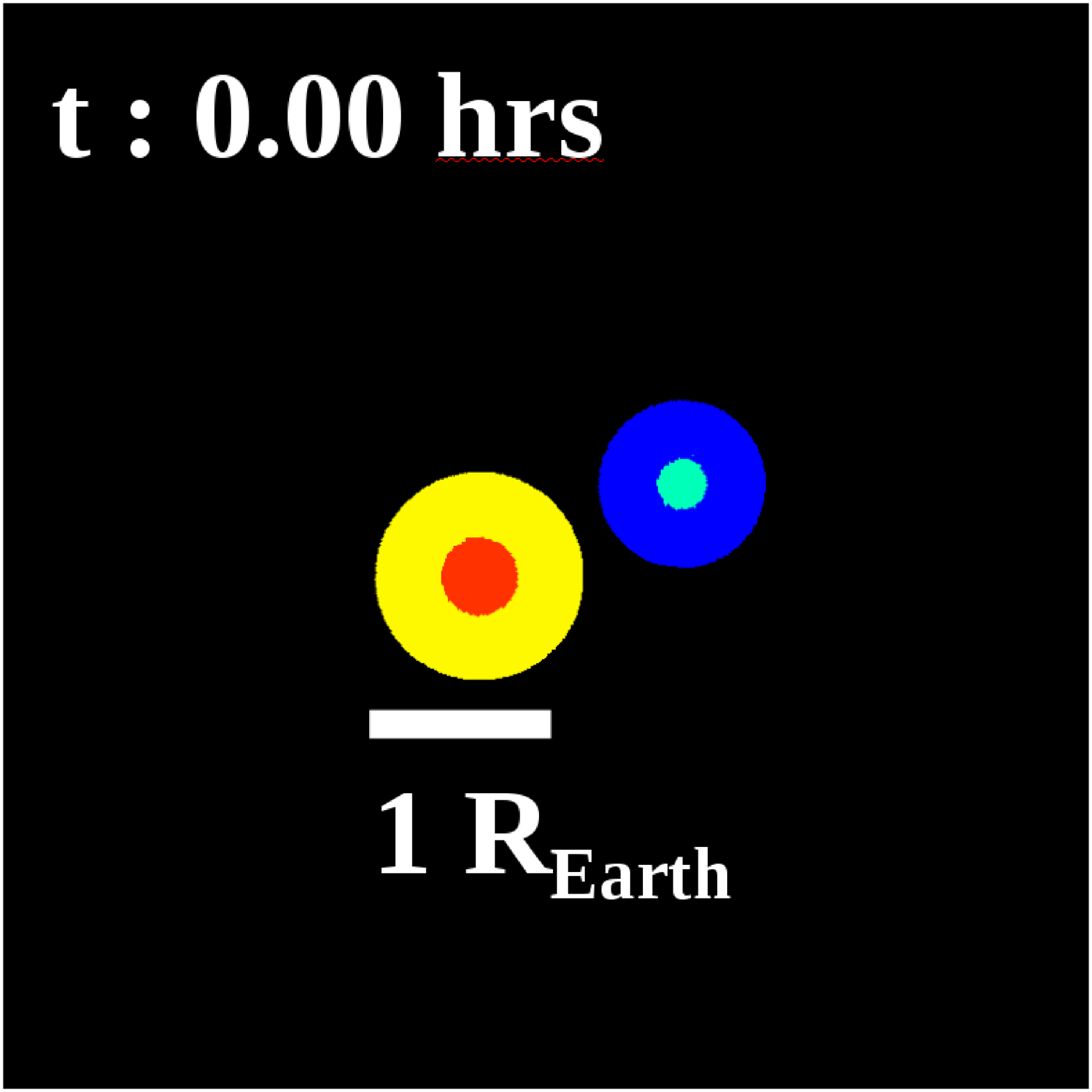}
\includegraphics[width=0.24\textwidth]{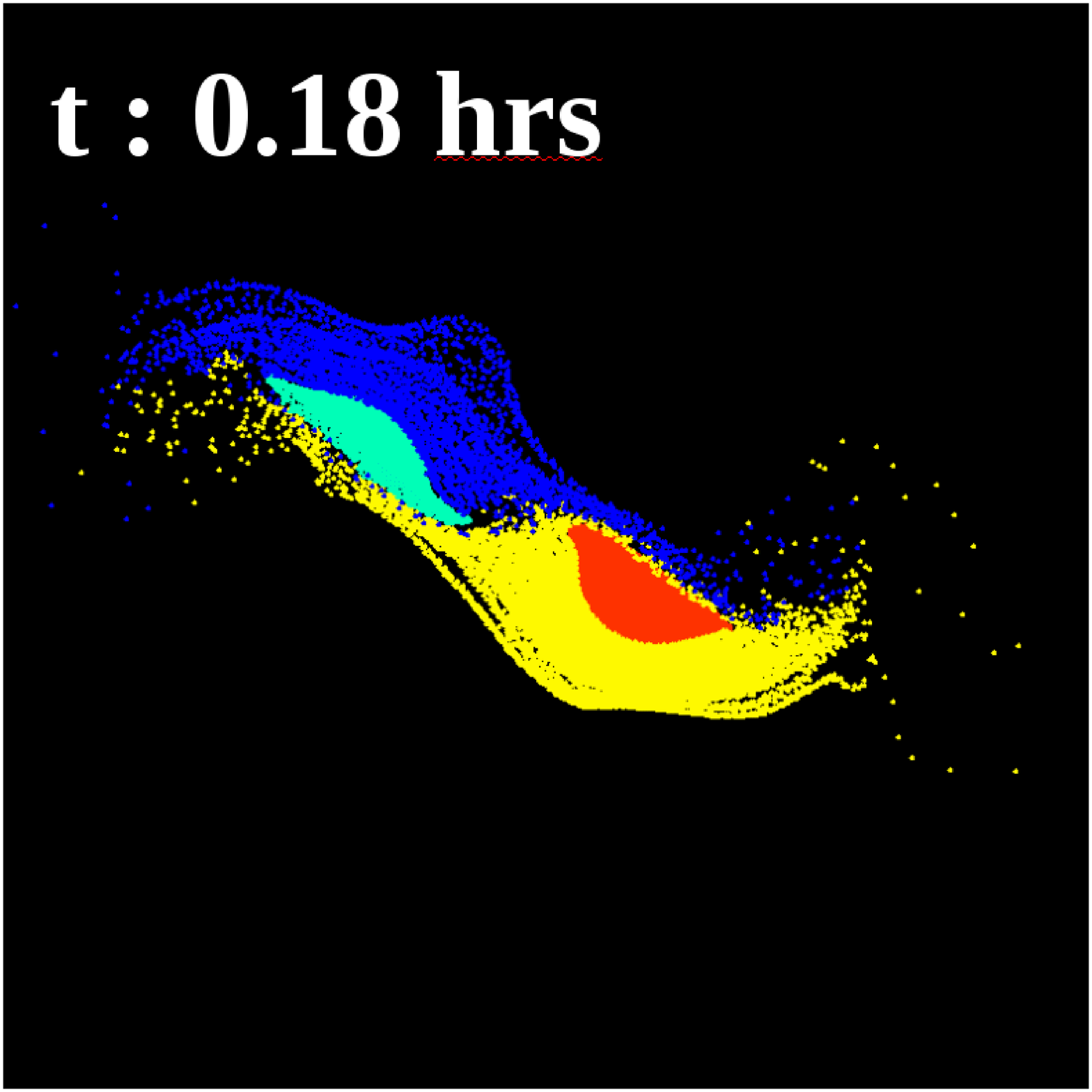}
\includegraphics[width=0.24\textwidth]{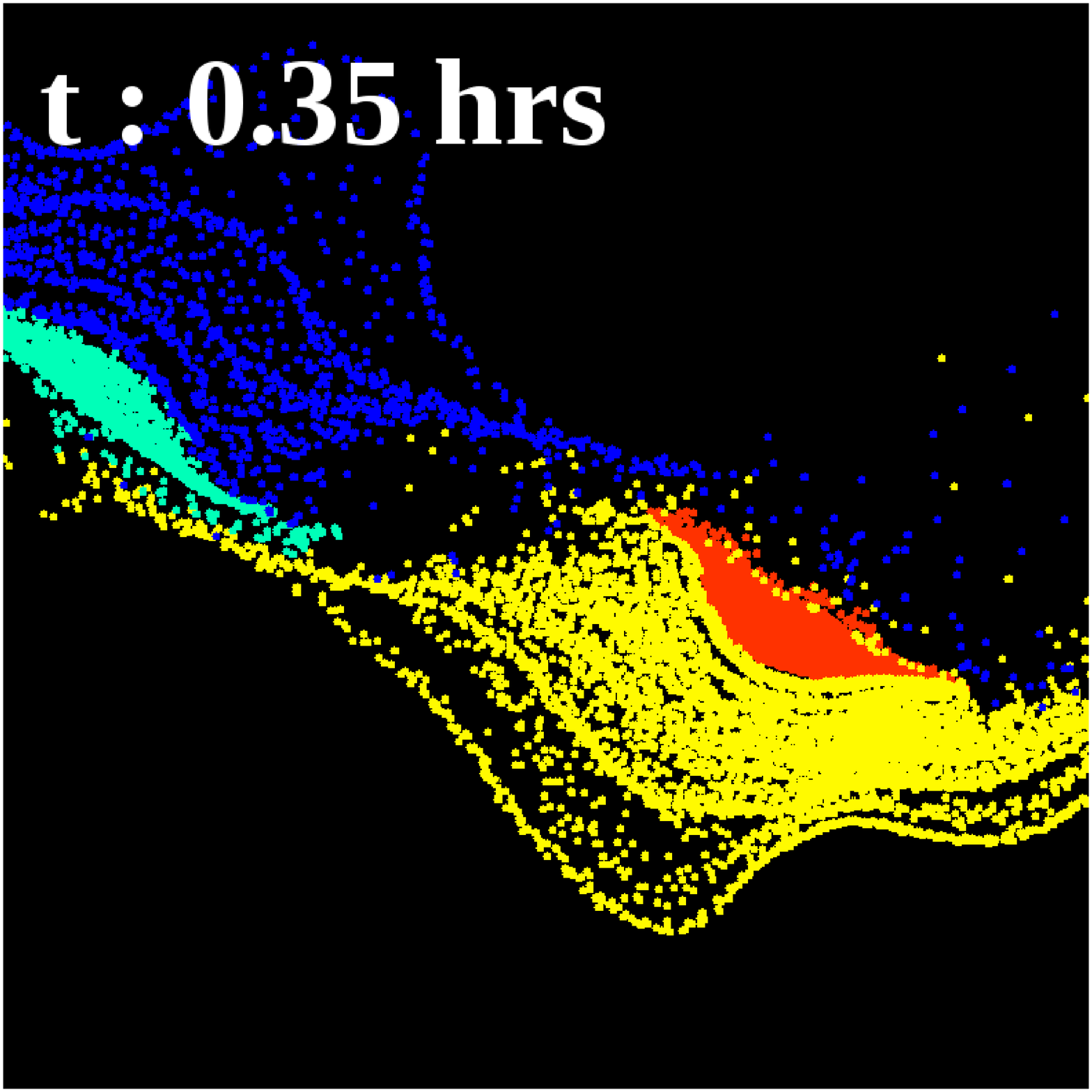}
\includegraphics[width=0.24\textwidth]{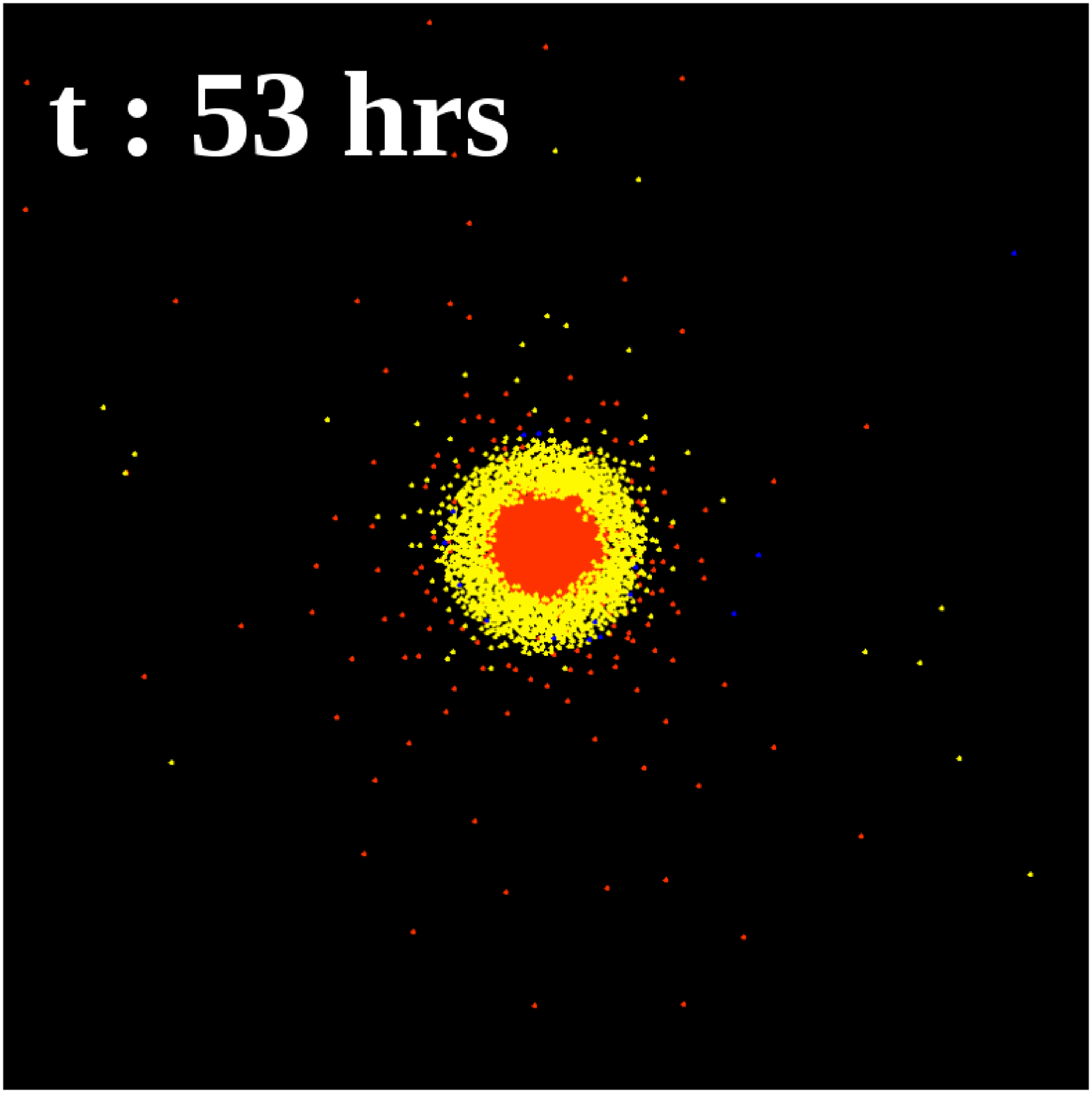}
\caption{Snapshots at different times of a Case-1 collision where the figure size is kept constant for the different times. The proto-Mercury of $2.25~$M$_{\mercury}$ (red: core, yellow: mantle) collides with the impactor of $1.125~$M$_{\mercury}$ (turquoise: core, blue: mantle) at an impact angle of $b=0.5$ and $v=30$~km/s. In the last frame, we show the largest fragment of 0.95 $M_{\mercury}$ and $Z_{Fe}$=0.67. { The snapshots are made in the x-y plane, and with z=[-0.2,0.2]~$R_{\oplus}$, and the bodies are seen from the top.}\label{Snapshot1}}
\end{center}
\end{figure}
\begin{figure}[h!]
\begin{center}
\includegraphics[width=0.24\textwidth]{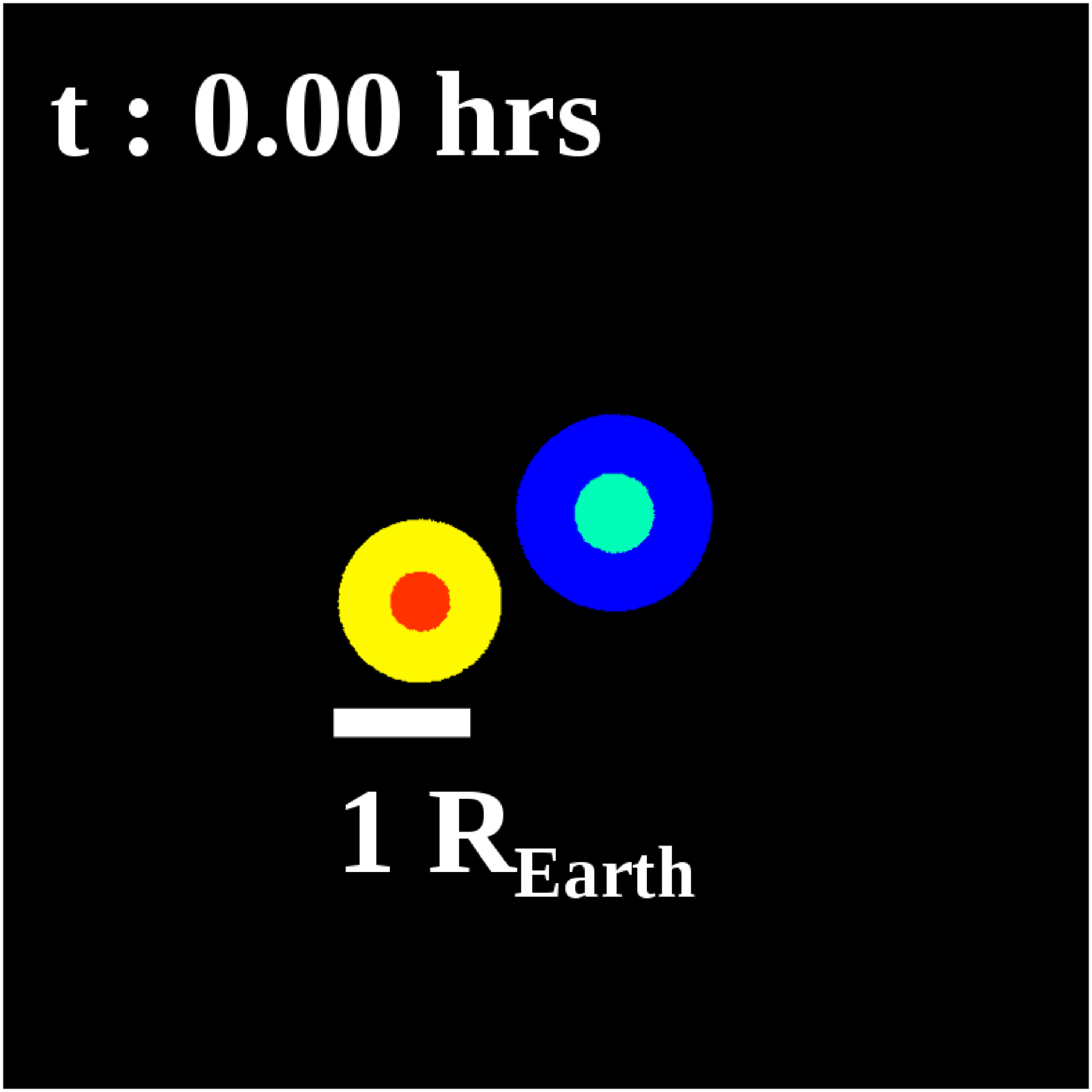}
\includegraphics[width=0.24\textwidth]{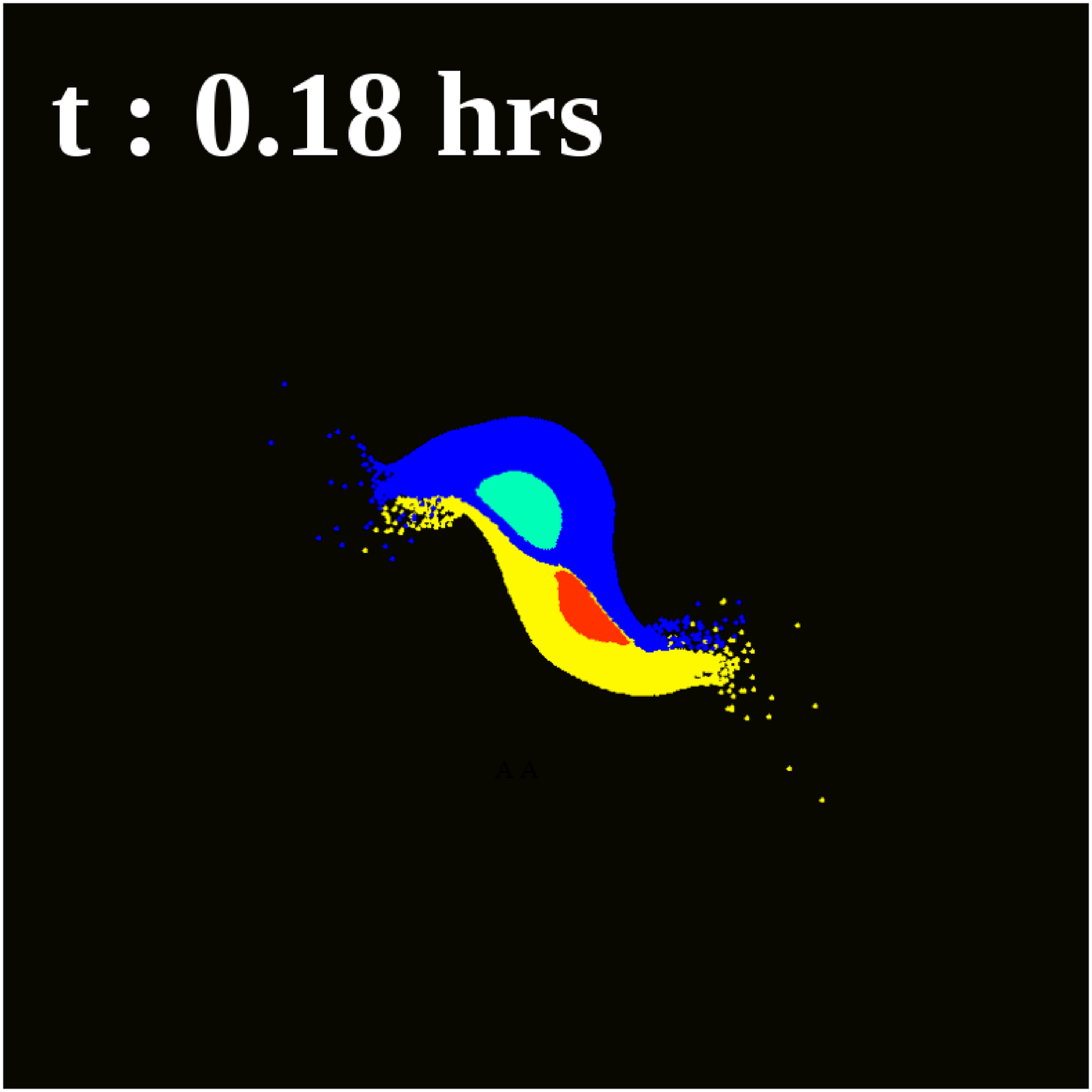}
\includegraphics[width=0.24\textwidth]{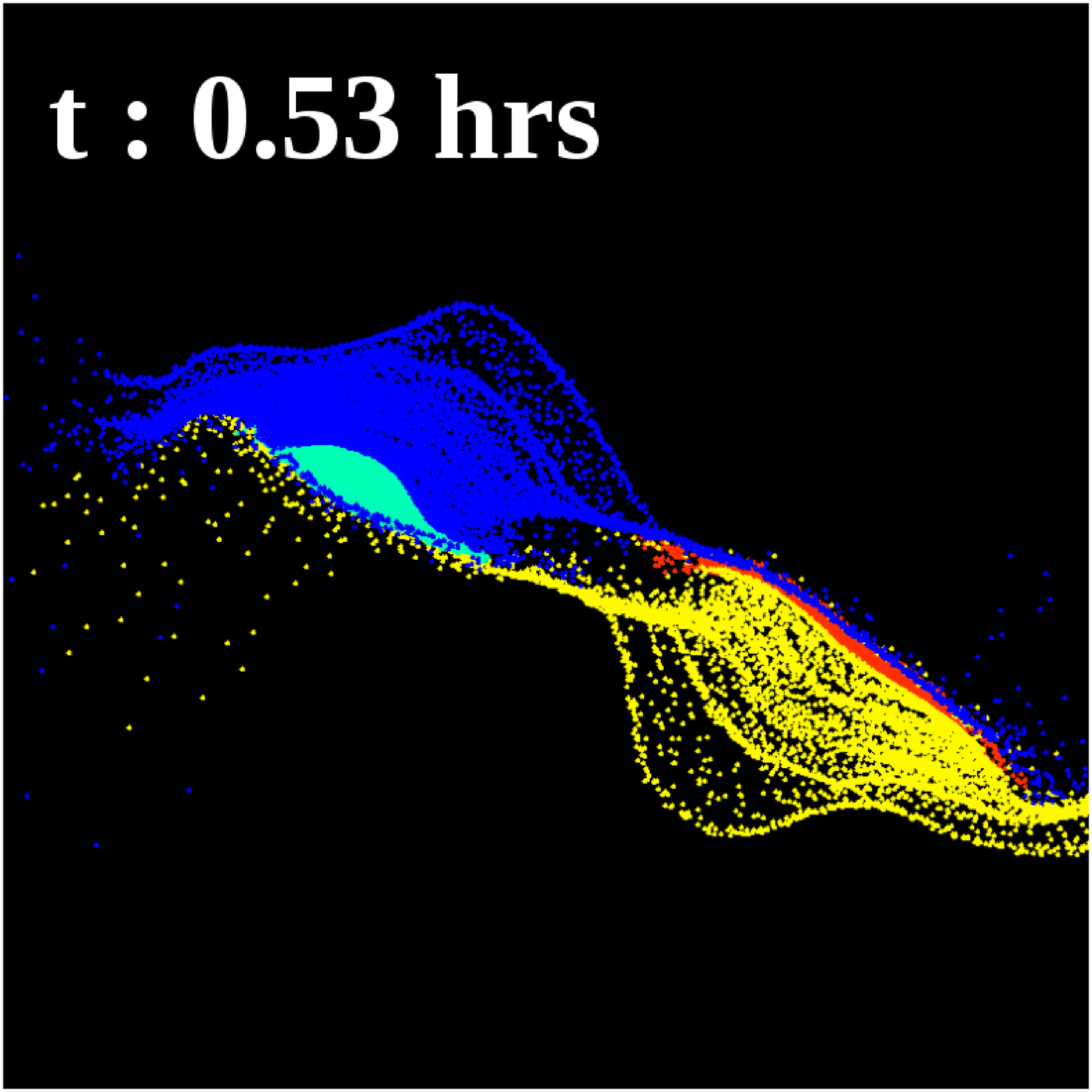}
\includegraphics[width=0.24\textwidth]{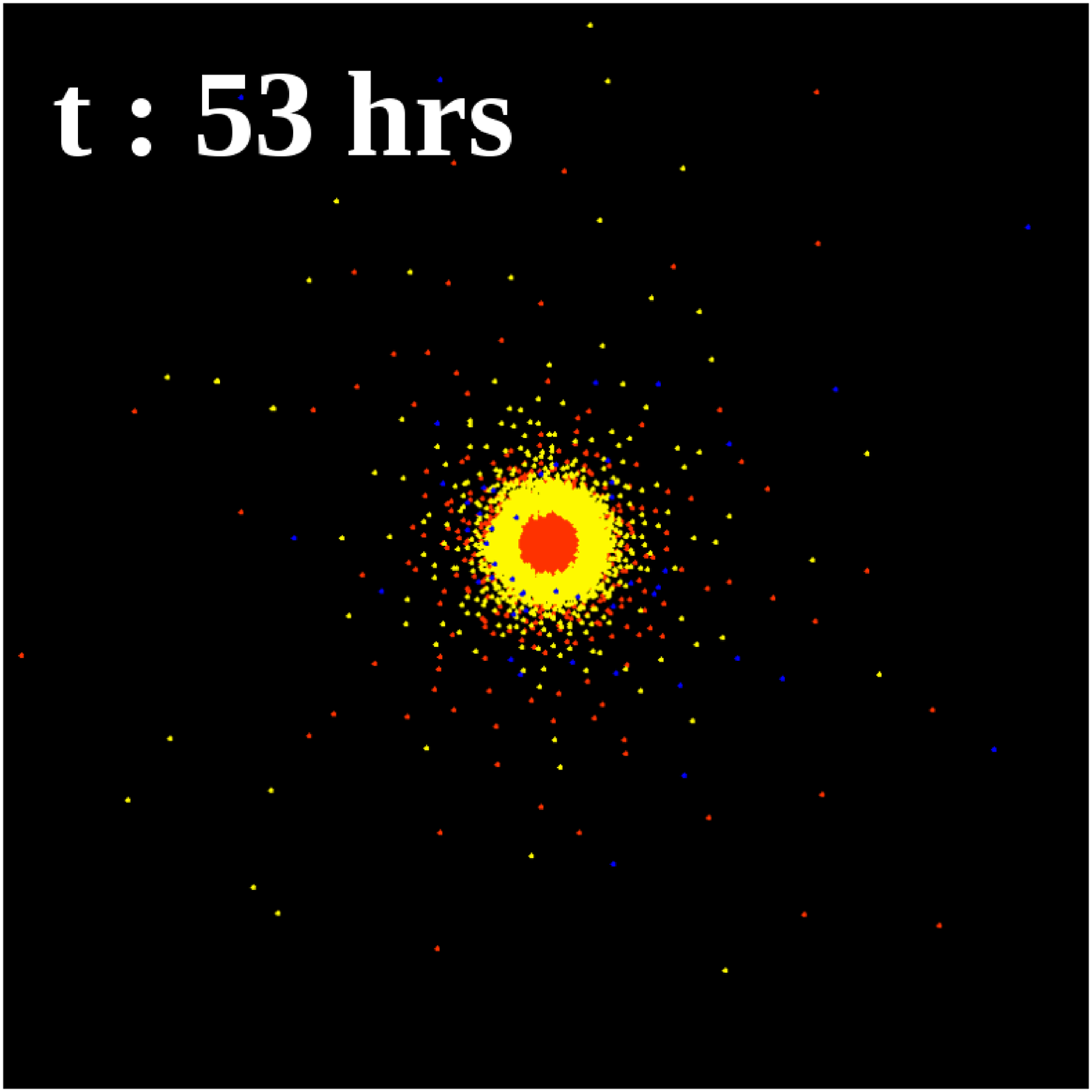}
\caption{Same as Figure 1 but for a Case-2 collision. The proto-Mercury of $2.475~$M$_{\mercury}$ collides with an impactor of $4.53~$M$_{\mercury}$ at an impact angle of $b=0.5$ and $v=20$~km/s. In the last frame, we show the second largest fragment of 1.08 $M_{\mercury}$ and $Z_{Fe}$=0.56. \label{Snapshot2}}
\end{center}
\end{figure}

\begin{figure}[h!]
\begin{center}
\includegraphics[width=0.5\textwidth]{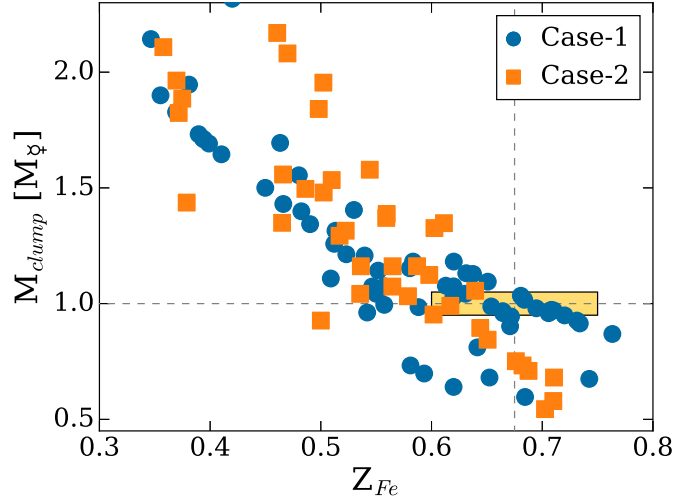}
\caption{The mass of the gravitational bound clump as a function of $Z_{Fe}$. Each point represents an outcome of a pair collisions: blue points stand for Mercury being the target and the largest remnant of collision while orange cases stand for Mercury being the impactor and second largest remnant of a collision. The mass is normalized to Mercury' mass and the gray lines indicate Mercury reference case. The yellow rectangle shows the region that is Mercury-like. \label{allresults}}
\end{center}
\end{figure}

Figures \ref{Snapshot1} and \ref{Snapshot2} present snapshots of our numerical simulations for Case-1 and Case-2, respectively. Figure \ref{allresults} shows the resulting iron mass fraction vs.~clump's mass for all the simulations we performed.  
We find that only a few cases lead to the desired region in mass and $Z_{Fe}$ as indicated by the yellow rectangle. 
Mercury-analogs are allowed to differ by at most 5\% from Mercury's current mass and have $Z_{Fe}$ between 60 and 75\%. 
The former is the scatter we allow while the latter is the uncertainty given by structure models with MESSENGER measurements of $C/MR^2$, the normalized moment of inertia, and $C_m/C$, the fraction of the polar moment of inertia contributed by the solid outer shell \citep{Hauck2013}\footnote{ $R$ is the planet`s radius, and the planet's mass is given by $M=4\pi \int_0^R \rho(x)x^2\mathrm{d}x$. $C$ is the polar moment of inertia 
$C=\frac{8\pi}{3}\int_0^R \rho(x)x^4\mathrm{d}x$, 
and $C_m$ is the one due to the solid outer shell defined by 
${C_m}/{C}+{C_c}/{C}=1$.}.
It should be noted that the apparent correlation is linked to our initial conditions where we start with a fixed proto-Mercury mass and composition and consider only cases that lead to stripping. 
The few cases that lead to merging or destruction of proto-Mercury are not considered. 
We conclude that forming Mercury from a giant impact is difficult since a Mercury-analog is produced only with some fine-tuning of the parameters. 
An analysis of the scaling-laws derived from all the simulations are presented in the Appendix \ref{appendix}.

\subsection{A Standard Giant Impact (Case-1)}
\label{GI1}

\begin{figure}[h!]
\begin{center}
\includegraphics[width=0.48\textwidth]{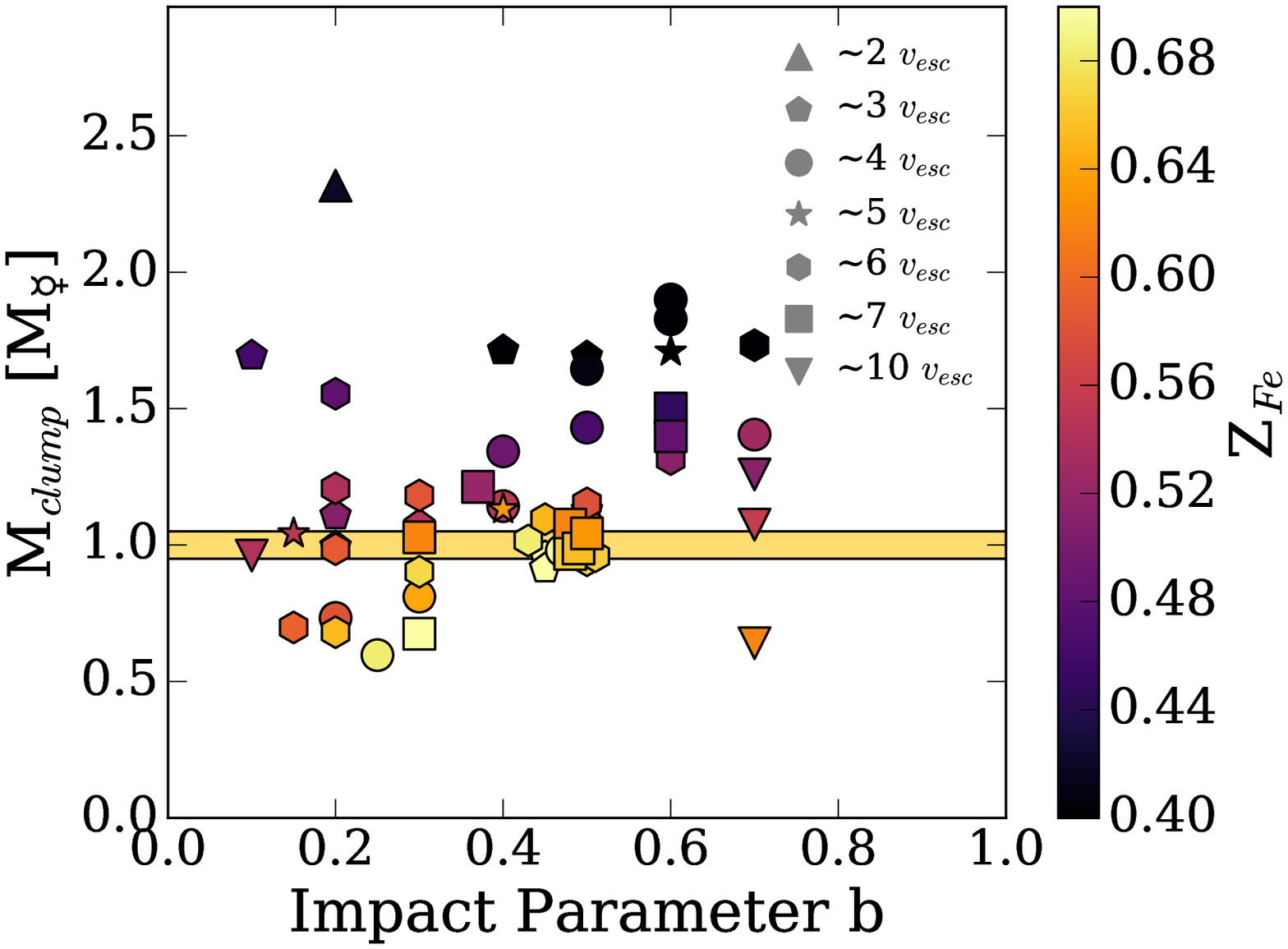}
\includegraphics[width=0.48\textwidth]{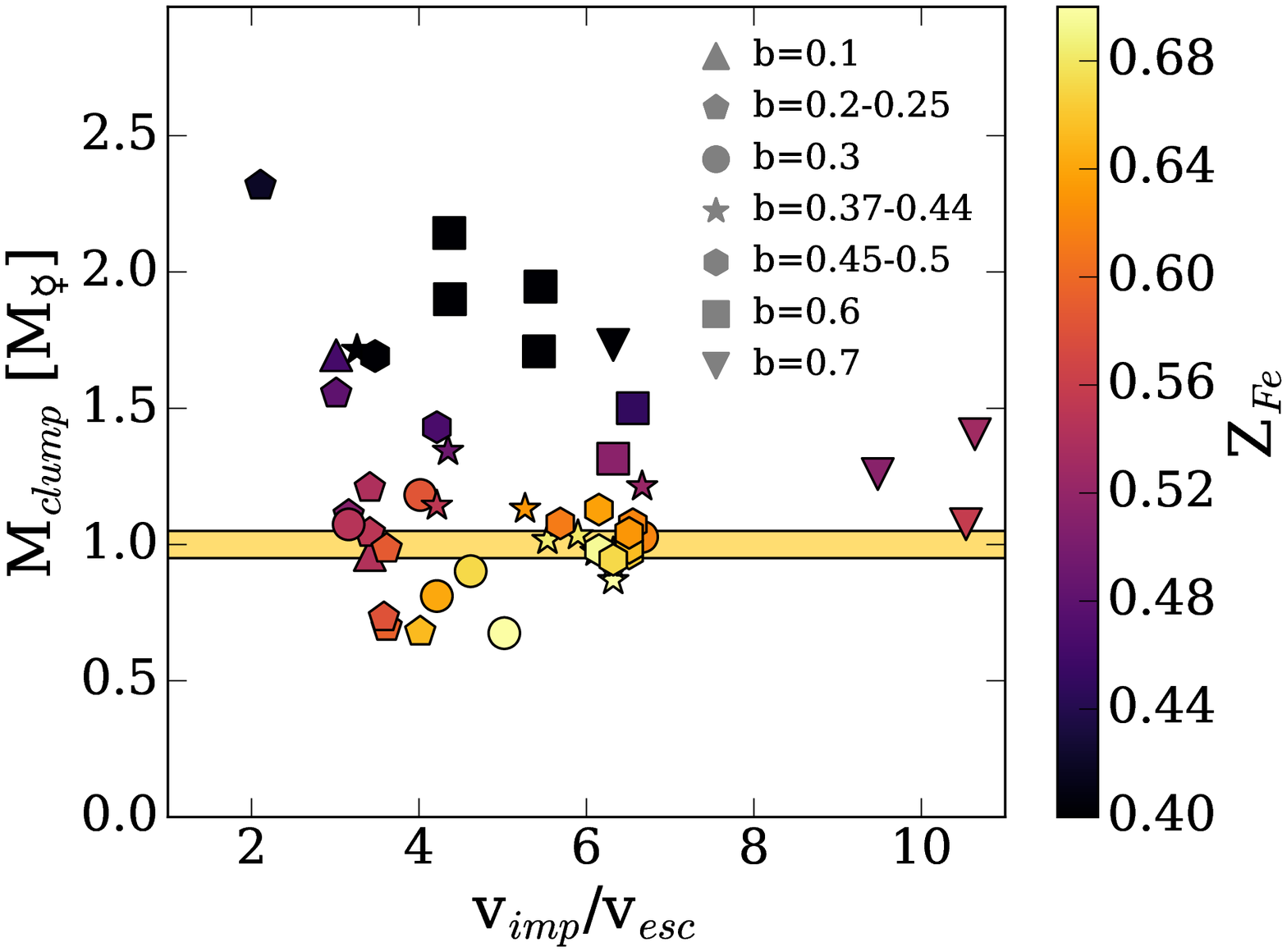}
\caption{The mass of the gravitational bound clump as a function of the impact parameter $b$ (left) and of the impact velocity in terms of the escape velocity (right) for Case-1. Each point represents the largest remnant mass of the pair-collision. The mass is normalized to Mercury's mass and the yellow band indicates the desired region in mass. The color map indicates the $Z_{Fe}$ value, with the promising results in dark-orange and red. \label{bmc}}
\end{center}
\end{figure}

Figure \ref{bmc} shows the final clump mass vs.~impact parameter and the escape velocity in Case-1. 
We notice a correlation between $b$, $M_{clump}$ and $Z_{Fe}$: the more head-on the collision is, the more mass is stripped leading to a higher final iron mass fraction. The collision chances are limited because of the small size of impactor: since the collisional cross-section is as large as the impactor, it limits the maximum energy that is transferred to the target.
For a given velocity, the outcome of the collision can be reduced to a geometry problem: more head-on collisions lead to deeper penetration of the impactor towards the target iron core, and more mantle stripping. 
Some of the iron in the core can be lost in such cases, requiring us to increase the initial target mass by 10 to 20\% in order to achieve the desired final $Z_{Fe}$.

To reproduce Mercury's properties, collisions must be quite head-on ($0.3 < b < 0.55$) with velocities at least five times the typical escape velocity of the system, or $v_{imp}\sim$30~km/s, which is notably high. 
However for very head-on collisions, lower velocities are required (3-4$~v_{esc}$, i.e., 20~km/s) to prevent destruction of proto-Mercury and lead to the right $Z_{Fe}$. 
For lower velocities ($\sim 2~v_{esc}$), the mantle can be stripped but the impactor is re-accreted by the target (not shown in Figure~\ref{bmc}).

\subsubsection{Sensitivity to the Impactor's Composition}
\begin{figure}[h!]
\begin{center}
\includegraphics[width=0.5\textwidth]{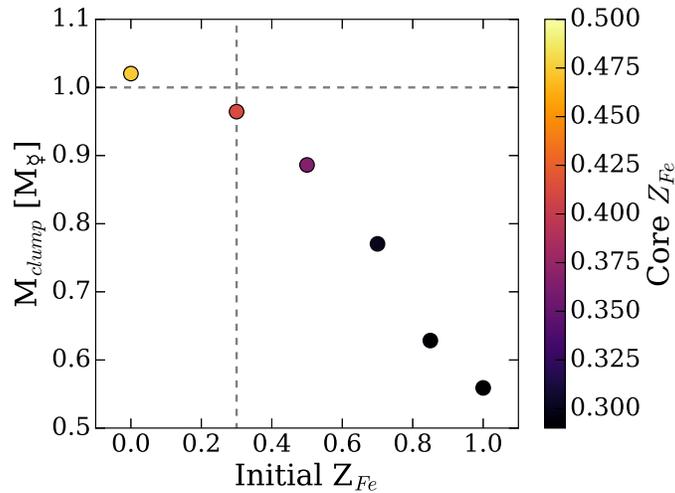}
\caption{ The sensitivity of the resulting target's mass and iron distribution to the impactor's composition. Shown is $M_{clump}$ vs. the initial $Z_{Fe}$. In all these cases proto-Mercury is hit by an impactor of 0.1~M$_{mpM}$ with various compositions (different initial $Z_{Fe}$). The colormap indicates the ratio of iron in the core, $M_{core, Fe}/M_{clump}$. The rest of the iron within the planet is distributed in the mantle and the vaporized material. The mass is normalized to Mercury's mass with the gray lines indicate Mercury's current mass value and the standard $Z_{Fe}$ of the impactor (30\%). \label{Compimp}}
\end{center}
\end{figure}

In the previous simulations, both the target and impactor were massive enough to be differentiated. 
We next consider a baseline case where proto-Mercury is hit by a small impactor (0.1 M$_{mpM}$) that could be non-differentiated. We consider impactors with various initial $Z_{Fe}$ as well as pure iron and pure rock impactors. 
We then investigate  the sensitivity of $M_{clump}$ and $Z_{Fe}$ to the impactor's composition. The impactor's mass is kept constant but the size changes in accordance to the assumed composition. The collisions occur with velocity $v_{imp}=7.5~v_{esc}$ (i.e., 30~km/s) and $b=0.3$. Figure \ref{Compimp} shows $M_{clump}$ and the core-to-clump ratio $M_{core, Fe}/M_{clump}$ vs. the initial $Z_{Fe}$, where  $M_{core,Fe}$ is the iron mass in the core. 
$Z_{Fe}$ is defined as the total iron mass fraction, and therefore does not discriminate between the condensed iron in the core and the liquid/gaseous iron in the atmosphere/mantle.

Interestingly, all the final clumps have the same mean $Z_{Fe}$ of 0.6 but the distribution of iron within the planet is different. 
The inferred $Z_{Fe}$ value is the same due to two competing effects:
(1) The denser the object is, the more it can penetrate through the planetary body and eject more material, both in the core and in the mantle. 
In our case the iron sphere induces a final clump that is lighter by 10\% than its differentiated counterpart. On the other hand, rocky impactors lead to final objects that are slightly heavier. 
Note that this effect is strictly different from one of geometry, where a smaller object at a smaller angle has a smaller impact surface and cannot strip as much material as a bigger object. 
Here we actually see the opposite effect. 
(2) Since the iron impactor can only contaminate the target with iron, it increases the final $Z_{Fe}$ while rocky impactors disrupt the iron core less, leaving proto-Mercury with a larger core, and can only contribute rock to the mantle. 

We find that the final planetary mass, the core's $Z_{Fe}$ (and the core mass fraction), depend on the impactor's composition. 
Denser impactors lead to smaller clump's mass and a smaller $Z_{Fe}$ in the core, i.e. more iron is present in the mantle and/or is vaporized. $M_{clump}$ and $Z_{Fe}$ in the core do not follow a well-defined function of the initial $Z_{Fe}$. 
Our inferred iron distributions correspond to times shortly after the collision and they are likely to change with time.
When the mantle is still in a magma state, the iron droplets in the outer parts are expected to settle to the core without leaving a signature on Mercury's surface (e.g., \citealt{MagmaReview}). 
Once the mantle solidifies, the iron droplets remain in the mantle (and even on the surface). 
Therefore, current-state observations of Mercury's  iron distribution cannot be used to discriminate among the different cases. 

\subsection{Hit-and-Run (Case-2)} 
\label{GI2}

\begin{figure}[h!]
\begin{center}
\includegraphics[width=0.48\textwidth]{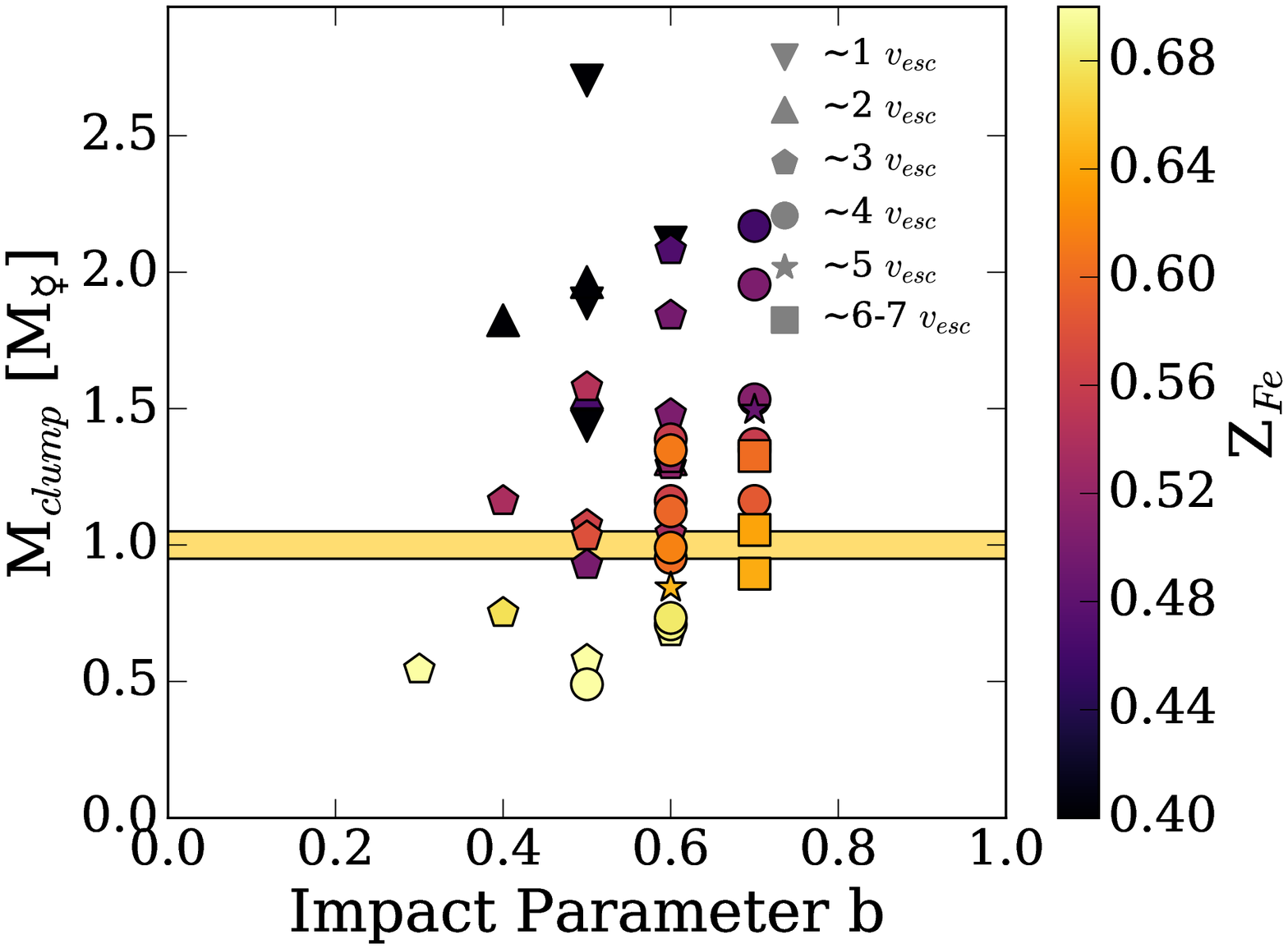}
\includegraphics[width=0.48\textwidth]{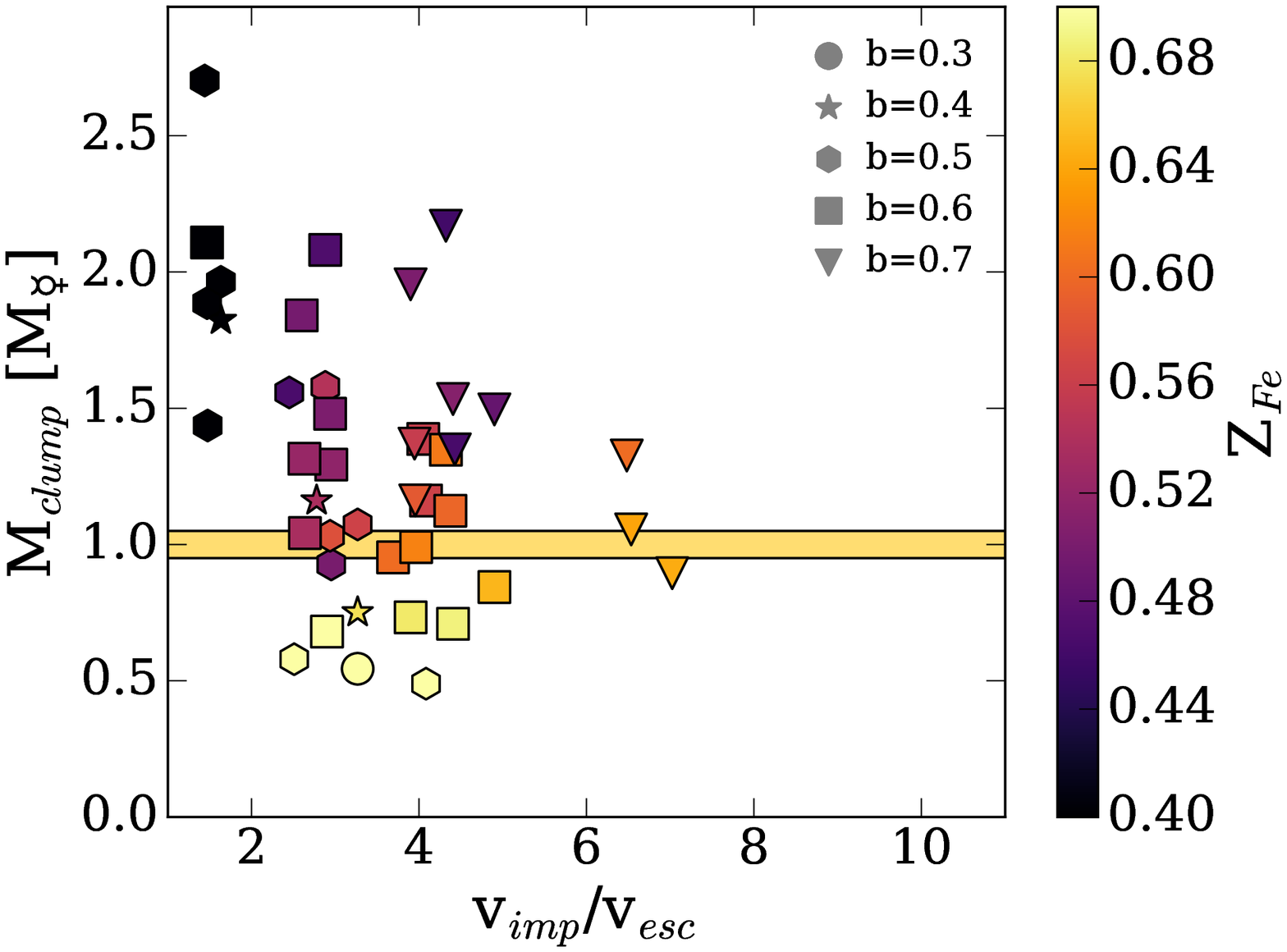}
\caption{The mass of the gravitational bound clump as a function of the impact parameter $b$ (left) and of the impact velocity in terms of the escape velocity (right) for Case-2. Each point represents the second largest remnant mass of the pair-collision. The mass is normalized to Mercury's mass and the yellow band indicates the desired region in mass. The color map indicates the $Z_{Fe}$ value, with the promising results in dark-orange and red.  \label{bmc2}}
\end{center}
\end{figure}

Figure \ref{bmc2} shows the final clump mass vs.~impact parameter and the escape velocity in Case-2.
In the hit-and-run scenario there are more possible combinations: a large $Z_{Fe}$ can be achieved with various impact parameters. However, we find that in these collisions more mass is removed from proto-Mercury than in Case-1. 
This is because in this configuration the interacting area and mass are larger. For a given $b$ the impactor transfers more energy than in Case-1 which affects the results. 
For example, collisions with $b=0.3$ and a velocity of ~20~km/s ($3~v_{esc}$) can lead to Mercury's $Z_{Fe}$, unlike in Case-1. This also holds for impacts with $b=0.6-0.65$ that are energetic enough (with $v_{imp}=30$~km/s, $\sim$ 4-5 $v_{esc}$). 
These energetic collisions can also entirely destroy proto-Mercury: in several cases, the collisions were found to be disruptive. 
Therefore we find that in Case-2 we must start with a larger initial mass of 3 to 5~M$_{\mercury}$ while we performed our simulations with initial masses of 2.475 to 3.375~M$_{\mercury}$. This is in agreement with \cite{Asphaug}'s most promising result with a proto-Mercury of 4.5~M$_{\mercury}$. 
We conclude that more configurations in Case-2 can lead to a Mercury-like planet. This case is also characterized by a lower ratio of $v_{imp}/v_{esc}$ than for Case-1. 
This is the consequence of using the same chosen velocity range as in the standard case with more massive systems, which leads to escape velocities higher than in Case-1.

\section{A Multiple-Collision Scenario \label{resultsmulti}}
A disadvantage of the single giant impact hypothesis is the relatively low probability of such violent collisions, and the required specific initial conditions. 
Although such giant impacts could occur during the final stages of terrestrial planet formation (e.g., \citealt{Quintana}), most collisions should occur with  an impact angle of 45$\degree (b=0.7)$  at mutual escape velocity \citep{Shoemaker}. 
In addition, the consistency of the giant impact hypothesis with MESSENGER's observations remains an open question. 
A giant impact is expected to affect a large part of the planet's mantle (and surface) via the formation of a magma ocean. Observational constraints for the existence of a magma ocean from models predictions, such as the surface elemental abundances, have not yet been presented and this requires detailed modelling of Mercury's 
post-impact evolution (e.g., \citealt{Gobalek}). A relatively high fraction of volatiles could be achieved by multiple collisions, in which each collision deposits less energy than a single giant impact, therefore leading to a smaller volatile depletion.
\par
It is therefore possible, and maybe even more probable, that Mercury formed as a result of multiple impacts.  
Considering multiple collisions enlarges the possible parameter space, and therefore we concentrate on several specific subsets. 
Below, we investigate multiple-collision scenarios that lead to Mercury's mass and $Z_{Fe}$. 

\subsection{Number of Impacts}
First, we study how many impacts are required to strip proto-Mercury's mantle. 
We consider two scenarios: (i) an angle of 45$^o$, i.e. $b$=0.7, the most probable angle, and an impact velocity of 3-4$~v_{esc}$, which is required to prevent mergers. 
The impactor's mass is 0.2, 0.3 and 0.5 M$_{mpM}$. (ii) impact angles of $b=0.5$, 0.7 and 0.8, with impact velocities of 3-4$~v_{esc}$. The impactor's mass is 0.5~M$_{mpM}$. 
For the subsequent collisions, the target is modeled with the updated $Z_{Fe}$ with the impactor always taken to be a new object. 

\begin{figure}[h!]
\begin{center}
\includegraphics[width=0.48\textwidth]{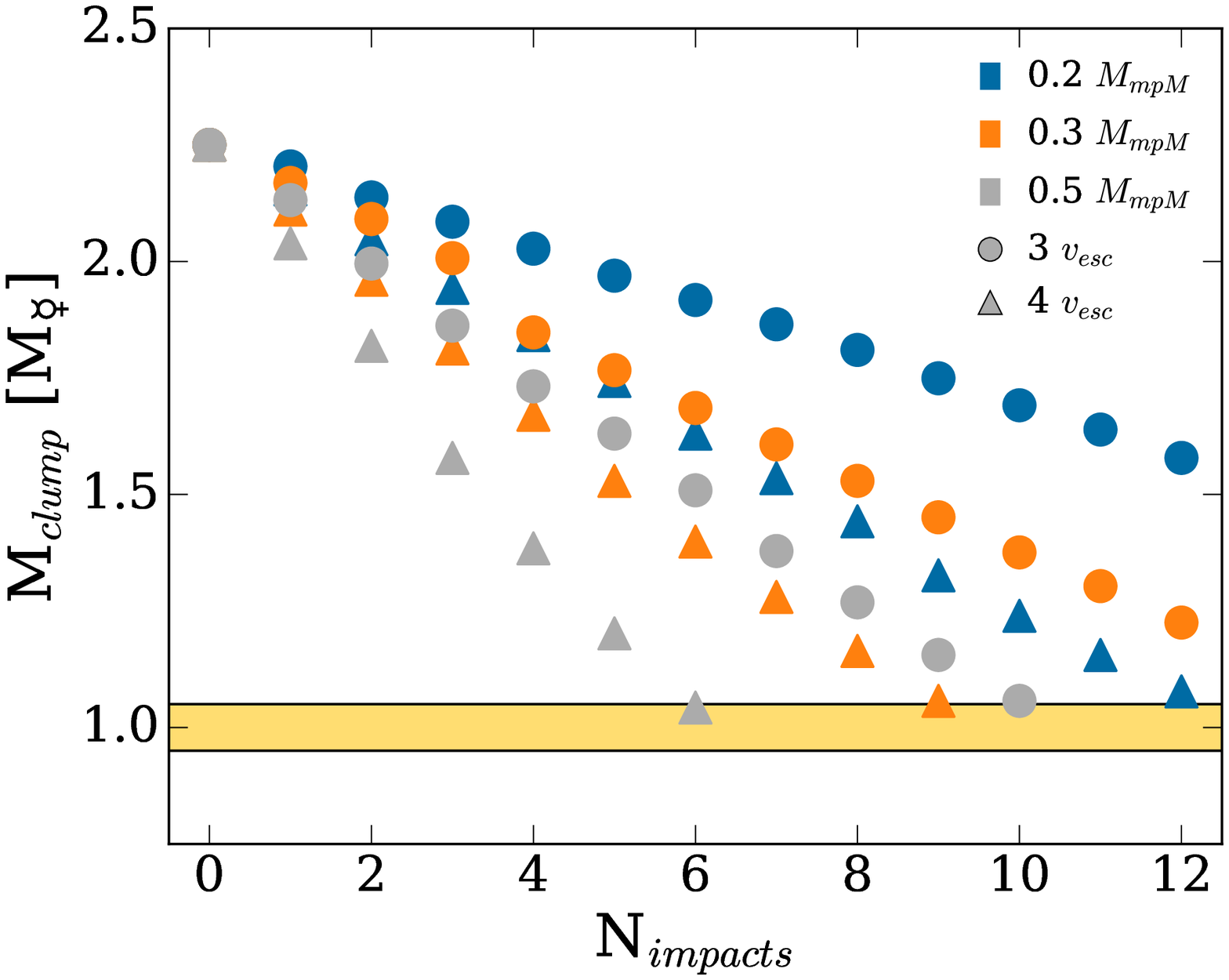}
\includegraphics[width=0.48\textwidth]{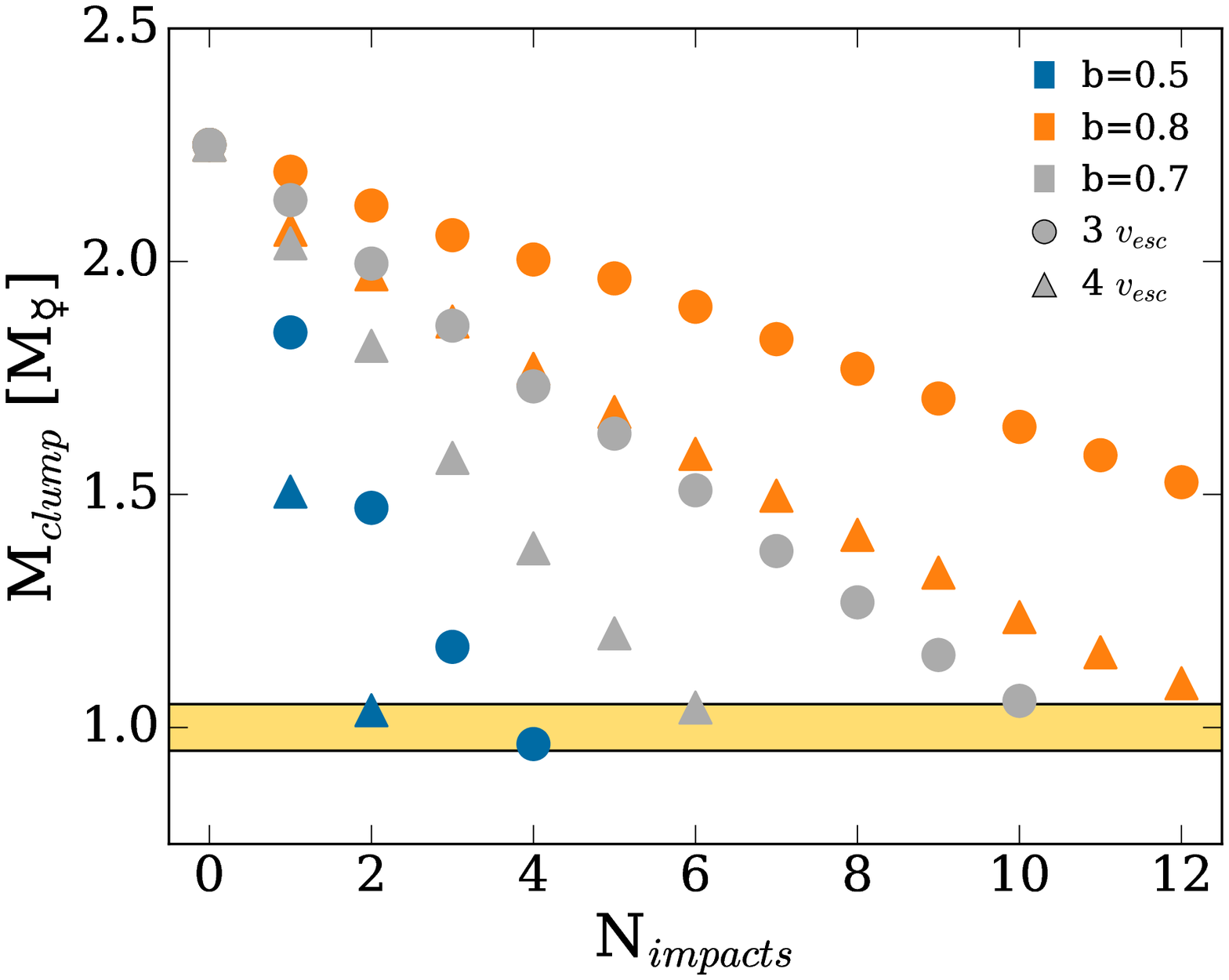}
\caption{$M_{clump}$ after each collision. The mass is normalized to Mercury' mass and the yellow band indicates the desired region in mass. \label{Sevimp}}
\end{center}
\end{figure}

Figure \ref{Sevimp} shows the mass loss from proto-Mercury after repeated collisions for the two scenarios.  
As expected, a more massive and faster impactor strips away more material. 
In scenario (i), Mercury can reach its current mass after six impacts with $v_{imp}\sim 4~v_{esc}$ or alternatively with ten impacts with $v_{imp}\sim 3~v_{esc}$ with the impactor's mass being 0.5~M$_{mpM}$.  
On the other hand, if the impactor has a mass of 0.2 M$_{mpM}$ more than a dozen impacts are needed, up to twenty in the least optimistic case. 
Even with an impactor with half the target's mass, the number of required impacts is high. 
We therefore conclude that this formation scenario for Mercury is rather unlikely. 

In the second scenario, Mercury can reach its current $Z_{Fe}$ after two impacts with $b=0.5$ and $v_{imp}\sim 4~v_{esc}$, or alternatively after four impacts with $v_{imp}\sim 3~v_{esc}$. 
We find that colliding an impactor of 0.5~M$_{mpM}$ at angle $b=0.8$ is equivalent to an impactor of 0.2 M$_{mpM}$ at angle $b = 0.7$. 

We conclude that Mercury could form as a result of several collisions with less extreme conditions than the giant impact of Case-1 or Case-2 presented in Sections \ref{GI1} and \ref{GI2}. The exact number of collisions depends on the impactor's mass, the collision angle, and impact velocity. It is more likely that Mercury formed via two collisions with $b\sim0.5$ at moderate velocities of $v\sim20~$km/s than via ten collisions at the most probable angle. 

\subsection{Timing and Relaxation between Impacts }
In the case of multiple collisions the timescale between collisions and the state of the target must be considered.
 A successive collision can occur at any time after the first one. 
Collisions that occur shortly after each other (i.e., tight in time) are characterized by a non-spherical and hot target while in the case of collisions that are well-separated in time the body is more likely to be relaxed and differentiated.
We therefore consider different thermal profiles, corresponding to different cooling times, to model the post-impact target. In some cases, the planet is still surrounded by a hot cloud of low density material. 
We begin with one collision and then explore how the assumed thermal state affects the subsequent collision outcomes. 
The first collision is between a proto-Mercury of $2.25$~M$_{\mercury}$ and a body of $1.125$~M$_{\mercury}$, at $b=0.7$ and $v=4~v_{esc}$ and the subsequent proto-Mercury has a mass of 1.97~M$_{\mercury}$. 
The second collision is with another body of 1.125~M$_{\mercury}$ at the same angle and velocity.

\begin{figure}[h!]
\begin{center}
\includegraphics[width=0.19\textwidth]{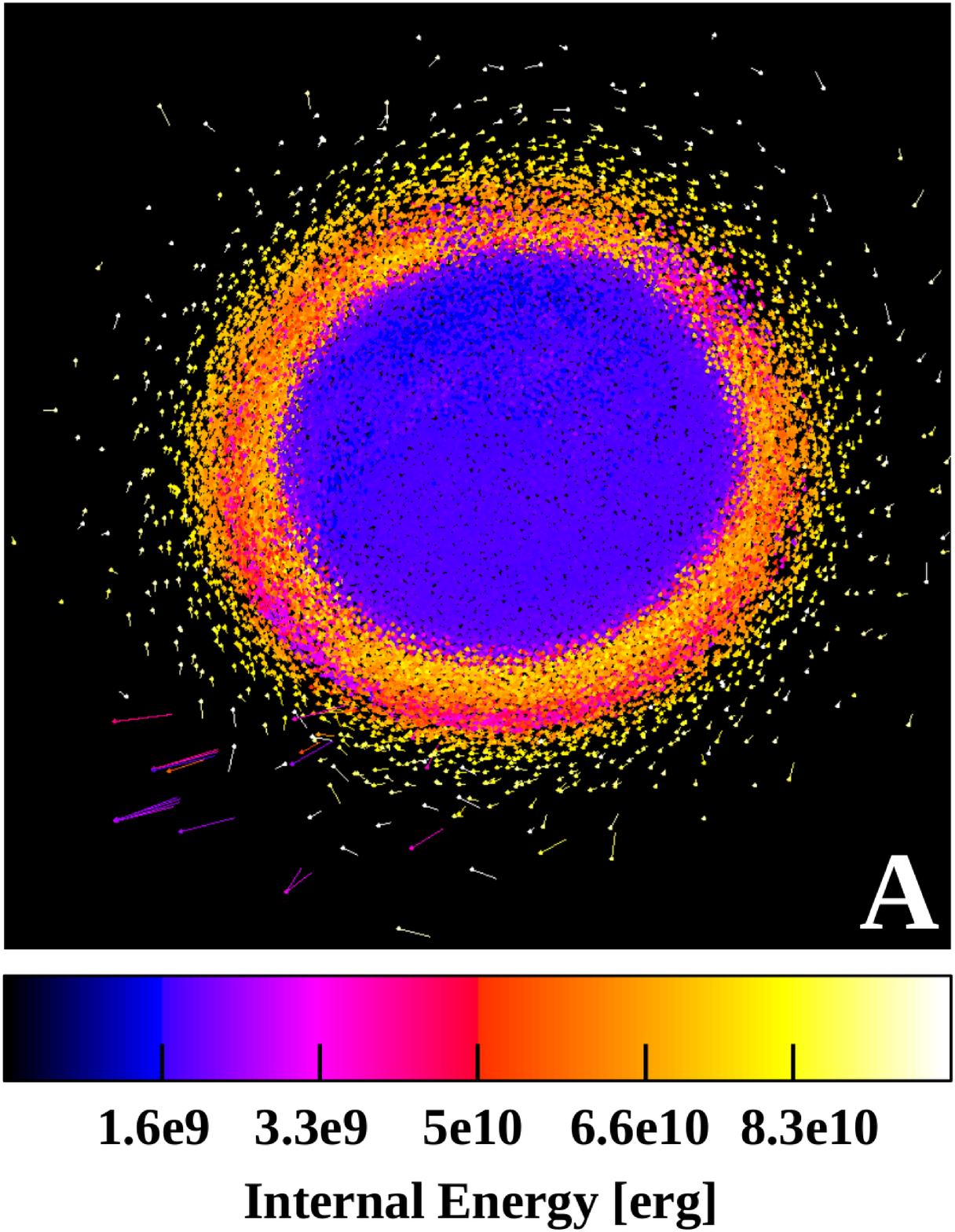}
\includegraphics[width=0.19\textwidth]{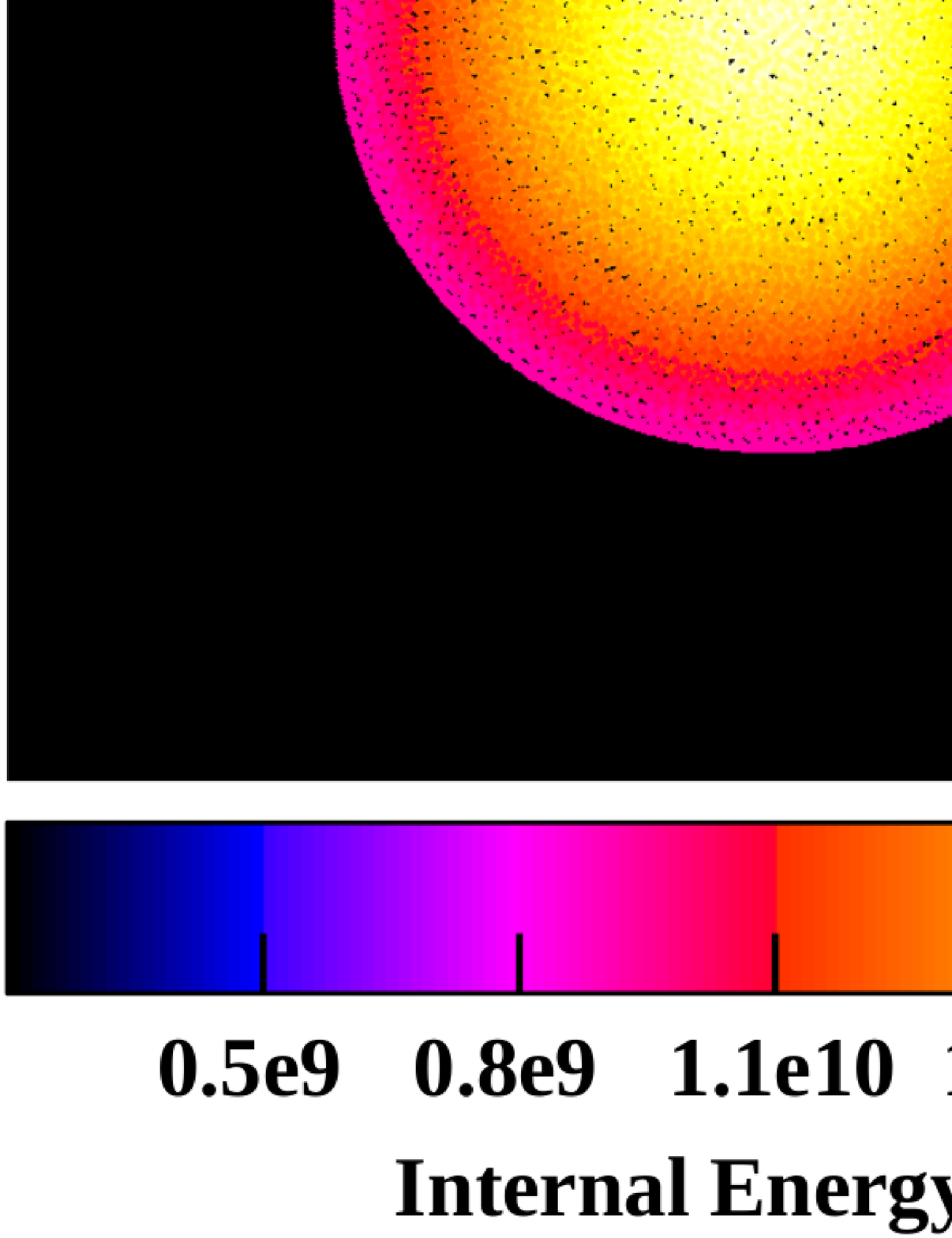}
\includegraphics[width=0.19\textwidth]{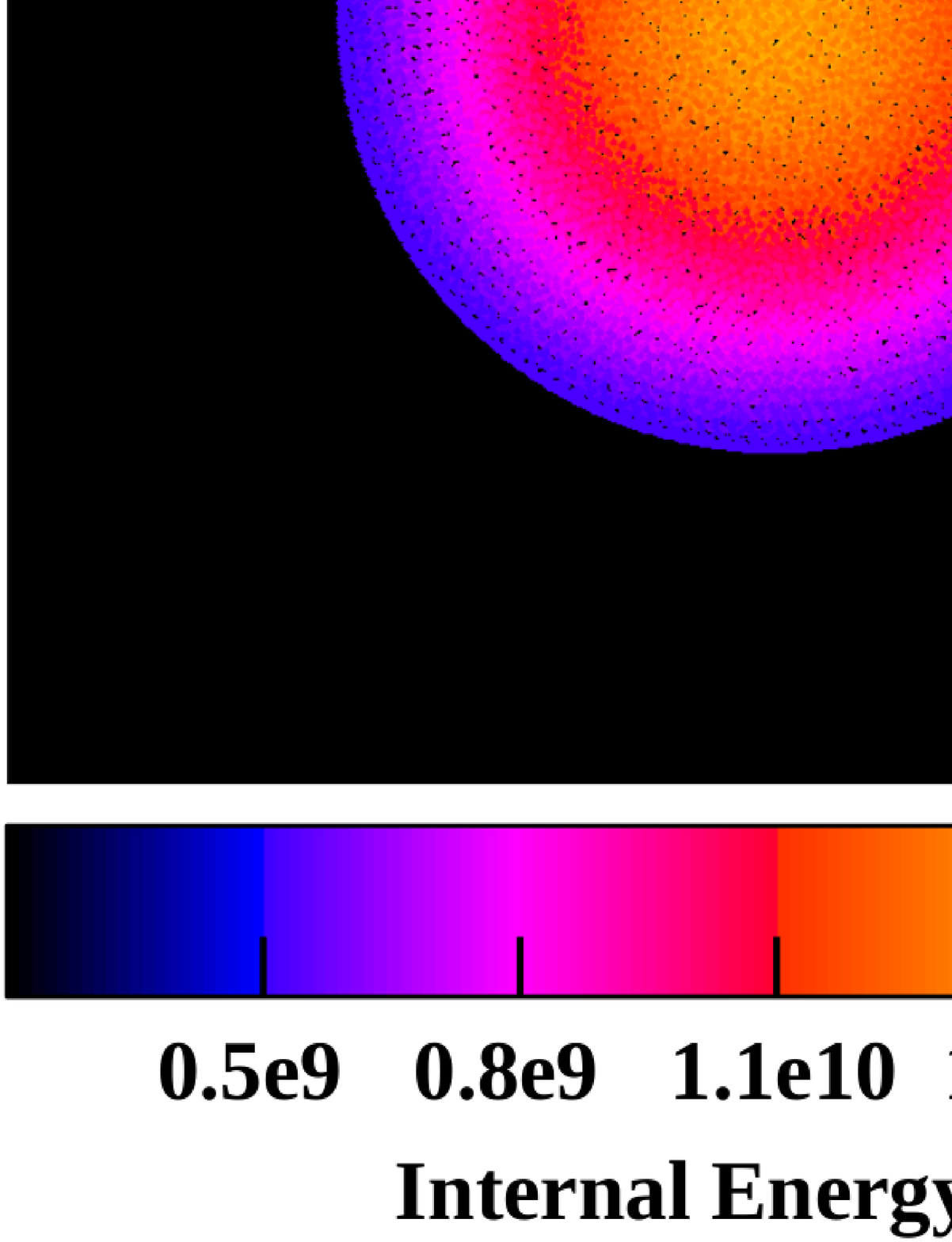}
\includegraphics[width=0.19\textwidth]{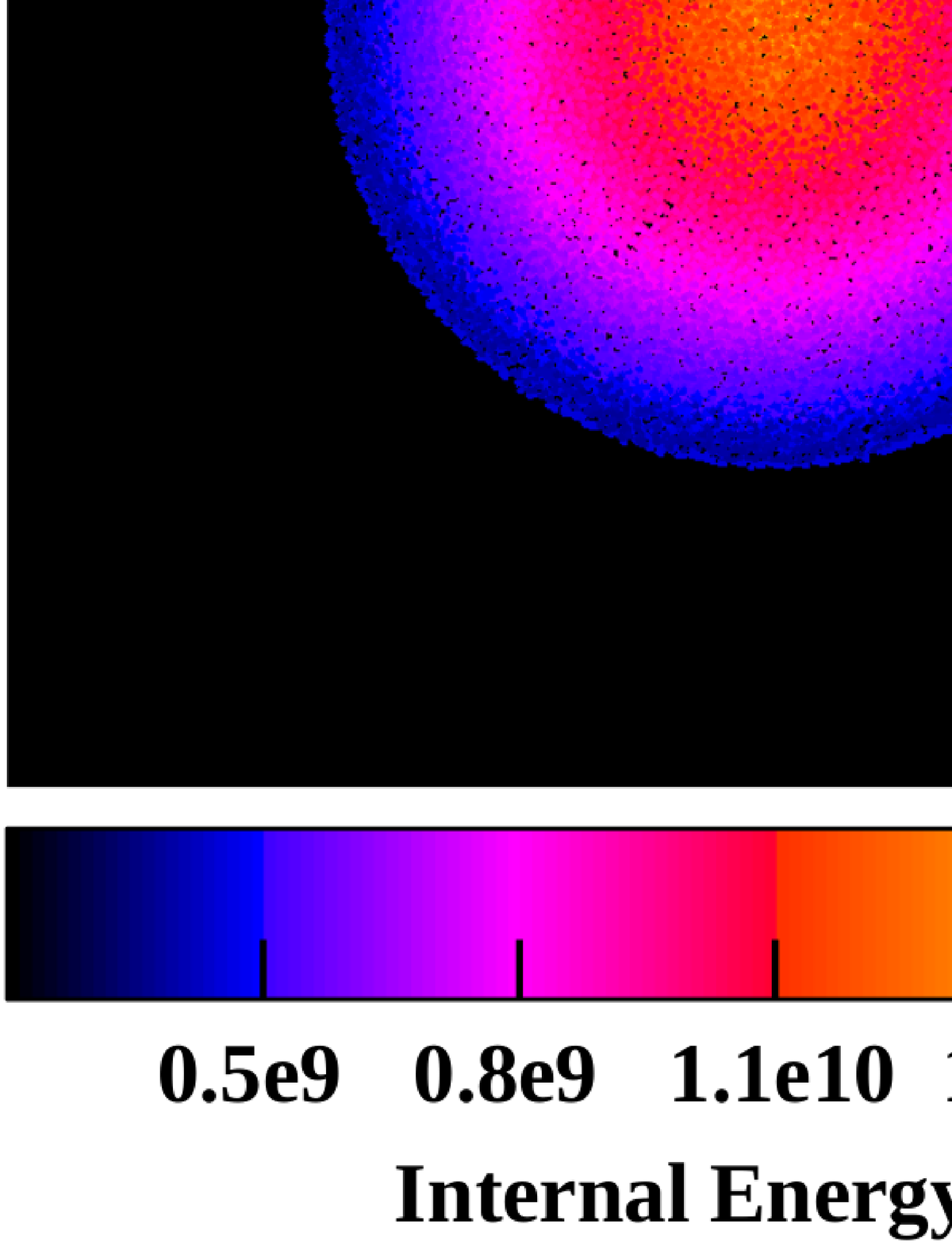}
\includegraphics[width=0.19\textwidth]{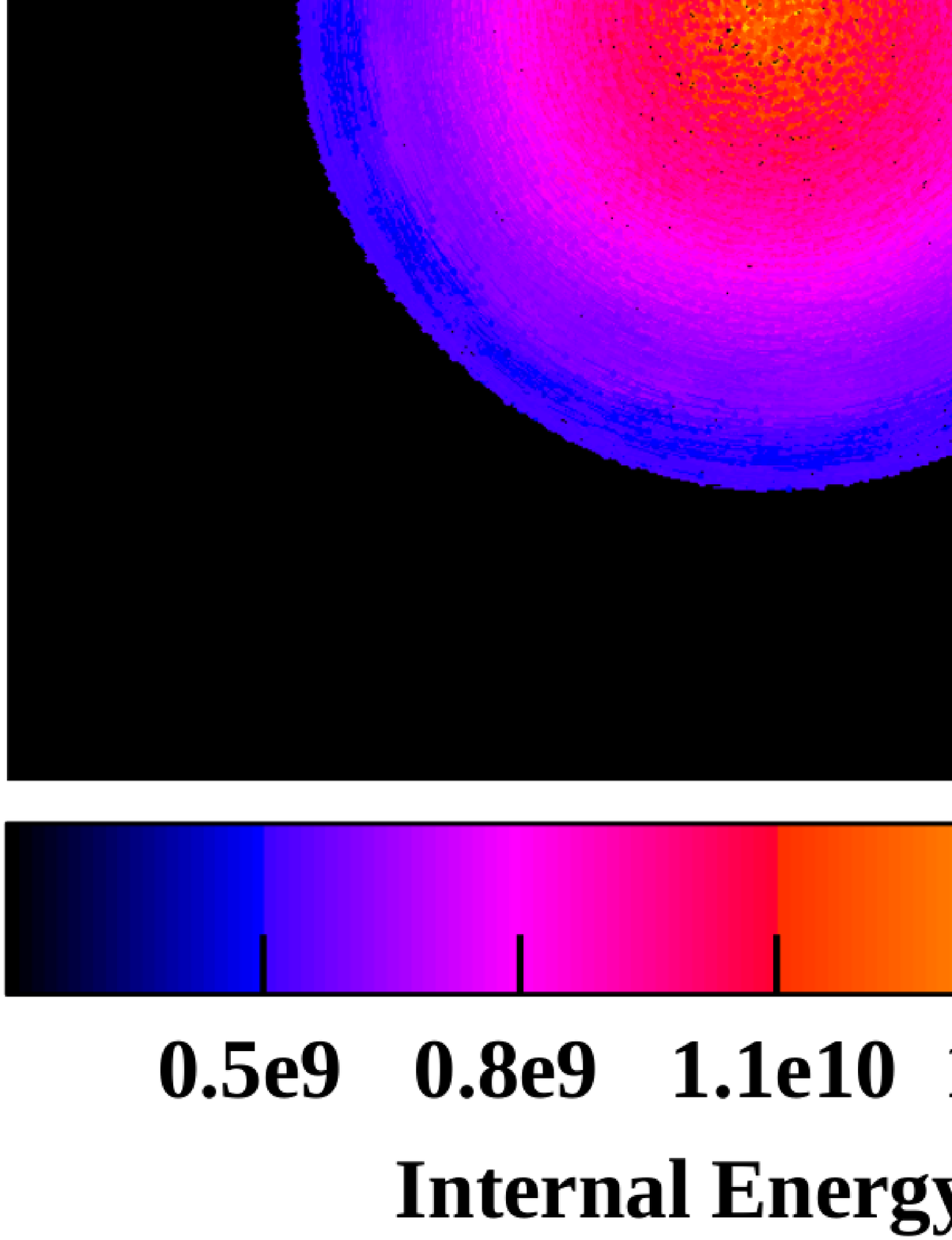}
\caption{Proto-Mercury with a mass 1.97~M$_{\mercury}$ for the different cases: (A) expanded (B) hot condensed (S) standard (non-rotating and cold) (C1) spinning T=30 hrs (C5) spinning T=5 hrs. 
The colormap corresponds to the internal energy value. Note that for case A the colormap has a different scaling. \label{SnapshotTarget}}
\end{center}
\end{figure}

The first case we consider (simulation setup A) is when the second collision occurs right after the first collision. 
Then, the object is still hot, mostly evaporated, and is non-spherical. This state corresponds to a timescale of $\sim$4-5 hours between the two collisions, i.e. 4-5 hours after the initial impact. 
While this scenario is somewhat unlikely as it requires nearly two simultaneous collisions, it is an interesting case to consider since it provides an upper bound for the envelope stripping in the multiple-collision scenario. 
We take the original simulation output to represent the target and use it directly in the subsequent collision. 
After the first collision, we include the particles that compose the gravitational-bound clump, with all stored quantities (e.g., density, internal energy, physical state of the material) and use it as an input for the second collision.  

In simulation setup B, we assume that the time between collisions is intermediate: the target is spherical as it has a moderate rotation period and its hot cloud has recondensed. 
The atmosphere is expected to cool down quickly by radiation. 
However if the surface of the body is still hot, the particles in the atmosphere could be heated and evaporate again. 
As a result, we estimate the cooling time as the time it takes the planet to radiate the energy associated with the impact 
and reach Mercury's equilibrium temperature of $\sim$600~K. With this simple consideration, which neglects the existence of an atmosphere, the cooling time is estimated to be $\sim$10$^4$~years.
This time estimate is appropriate until the rheological transition from the liquid to solid phase is achieved, which is $\sim$ 1300~K for silicates. 
A typical timescale for the surface temperature to drop below the melting temperature for 1/2 the Earth's mass is $10^3-10^4$ years \citep{Bonati2018, Bower2017}.
To account for this effect, we use a condensed-state structure but we increase the internal energy such that the surface temperature is $\sim$ 1200~K. 
Although the Tillotson EOS is not explicitly temperature-dependent, we can estimate $T_s$ using  the internal energy $u$. 
As there is no pressure, the contribution from the material molecular interactions also drops, which leaves $u=u_{thermal}=c_v T$. 
Assuming that the heat capacity $c_v$ does not change between 500~K and 1200~K, we can estimate $T_s$ and derive the new structure of the target. 

The last case we consider, simulation setup C, is when the target has cooled down completely between two collisions. 
In this case the target has a surface temperature of a few hundred K, near the equilibrium temperature assuming there is no significant internal heat source or mechanism to retain the heat. 
Evolution models suggest that a terrestrial planet could reach thermal equilibrium after billions of years, or geological timescales (e.g. \citealt{Lissauer}, \citealt{Stamenkovic}). 
For this case, we use the standard way to describe the target.
However, since the target has acquired some angular momentum in the collision we must consider its spin. 
To obtain the target's post-impact rotation period we bin the angular momentum of the SPH particles as a function of radius and determine the angular velocity by fitting a solid-body rotation to the data. 
The outermost layers are represented by only a few particles are noise-dominated and are therefore excluded.
Under these assumptions, we get a period of $P\sim$ 30~hours. This is a lower-bound since the outermost layers can contribute their (large) angular momentum as the target cools down. Possible deviation from solid-body rotation introduces an additional challenge in determining the rotation period accurately. For comparison, we also consider an intermediate and fast spinning body of $P \sim $ 10 and 5~hours (the break-up speed of a planet of Mercury's density is $P\sim$~1.5~hours). 
We then consider, for the same absolute value of angular momentum, both prograde (C1, C3. C5) and retrograde (C2, C4, C6) spins. 
We assume that the subsequent collision occurs in the same plane as the previous one.

When we consider the rotation acquired, the mass difference to the standard (non-rotating) case is about 1\%. 
We also cannot distinguish the retrograde and anti-retrograde cases. Overall this effect is small because the spinning velocities, respectively 1.2, 0.6 and 0.2~km/s, are much smaller than the impact velocity $\sim17.4$~km/s, 
and do not contribute much angular momentum to the system. 
We therefore conclude that mantle stripping is relatively insensitive to the target's rotation period in the high velocity collision regime, as the impact velocity plays a more critical role. We expect this effect to be profound in the low-velocity 
regime, where the impact velocity is of the order of the spinning velocity.
To summarize, we find that the multiple-collision scenario has different outcomes depending on the time interval between subsequent collisions. 
If the target has enough time to cool and re-condense, less material can be stripped away. 
For a given collision, up to 23\% of the planet's mass can be lost when the planet is still hot compared to 13\% for a compact and cold target. 
This indicates that the required number of impacts can be reduced if they are tight in time. 
Table \ref{lossmass} summarizes the different cases for the multiple-collision scenario. 

\begin{table}
\begin{center}
\begin{tabular}{l | l | l | l | l || l}
Setup & State & $t_{cool}$ [yrs]  & $R_p$ [R$_{\oplus}$] & $j_{tot}$ [$10^{12}$~cm$^2$/s] & $M$ [M$_{\mercury}$] \\ \hline
A & Expanded & 10$^{-5}$ & 0.65-0.75 & 4.3 & 1.25 \\
B & Condensed, $T_s=1200$~K & $10^4$ & 0.58 & 0.0 & 1.59 \\
S & Condensed, $T_s=600$~K & $10^9$ & 0.57 & 0.0 & 1.65\\
C1 & Condensed, P=30~hr & $10^9$ & 0.58 &  3.0 & 1.67 \\
C2 & Condensed, P=-30~hr & $10^9$ & 0.58 & 3.0 & 1.68 \\ 
C3 & Condensed, P=10~hr & $10^9$ & 0.57-0.59 & 9.1 & 1.68 \\ 
C4 & Condensed, P=-10hr & $10^9$ & 0.57-0.59 & 9.1 & 1.67 \\ 
C5 & Condensed, P=5~hr & $10^9$ & 0.53-0.62 & 18.2 & 1.67 \\ 
C6 & Condensed, P=-5~hr & $10^9$ & 0.53-0.62 & 18.2 & 1.67 \\ 
\end{tabular}\caption{Mass of the Mercury-like fragment after the second collision at different cooling times, corresponding to different initial states. After the first collision, the proto-Mercury \textbf{has} a mass of 1.97~M$_{\mercury}$. All the impactors \textbf{have a mass of} 1.125~M$_{\mercury}$ and collide at impact angle $b=0.7$ and $v_{imp}$=4~$v_{esc}$ ($\sim$ 17 km/s). The standard simulation set-up (S) is also shown for comparison, with a non-rotating target (P=0) amd a surface temperature of $\sim$ 600~K. \label{lossmass}}
\end{center}
\end{table}

\subsubsection{Impact Timing}

Another important aspect is the timing of the possible impacts. 
Proto-Mercury with a standard composition is expected to acquire its mass ($0.12$~M$_\oplus$) during the first 10-30 Myrs,
but can accrete $\sim$0.2~M$_\oplus$ within 100 Myrs \citep{Lykawka}.
If proto-Mercury collided with another impactor during the first 100 Myrs, mass can still be accreted, and at least another significant impact has to occur, supporting the multiple-collisions scenario. 
After 100 Myrs, as the proto-Mercury has accreted all of its mass, only one giant impact is required but several impacts are also possible. 
The latest impact has to occur within 1 Gyr, which is consistent with the upper bound on the Moon's formation via giant impact \citep{Quintana}. 
\par

The volatile-rich composition also gives constraints on the impact timing. 
A planet with a magma ocean has its volatiles diffused outwards during crystallization; these incompatible elements may be preferentially lost in the hit-and-run scenario \citep{Nittler2017}. 
The impacts we simulate here are energetic enough to melt at least half of the planet's mass \citep{Tonks}. 
This is consistent with either a single giant impact or with multiple impacts before the crystallizing of the magma ocean. 
The timescale of the latter is estimated to be $10^5$-$10^6$ years \citep{Elkins-Tanon}, which is very short. 
Overall, the impact timing is more constraining for the multiple collision scenario, especially given Mercury's volatile-rich surface composition. However, it cannot be used to discriminate among the different formation scenarios. 

\newpage 
\section{Discussion \label{discussion}}
\subsection{Ejecta Evolution \& Fate of the Impactor}
An important aspect in giant impacts is the evolution of the ejected material after the collision. The projectile deposits up to 70\% of its kinetic energy on the target (e.g., \citealt{Lissauer}). 
It is large enough to melt large parts of the targeted body: typically the kinetic energy is of the order a few times the required energy to melt and evaporate a basalt sphere of the target's mass. 
In such conditions, the mantle is expected to be in a magma phase while the ejected material is mostly heated and vaporized (e.g. \citealt{Tonks} for quantitative modeling). 
After the collision the material can cool down and re-condense directly or undergo the droplet regime before becoming solid again. Based on this simple thermodynamical argument, \cite{Benz2007} showed that the droplets or solid fragments are of the mm-cm size and are therefore subject to the Poiynting-Robertson drag. This effect, however, is expected to remove the particles only after $\sim$ 2.5 Myr, with 35-40\% of particles being reaccreted by Mercury. \cite{Gladman2009} suggested that the ejecta material would spread in a thin ring which is optically thick around the Sun that would reduce the drag. They estimated that the opacity can be larger than one in some cases. Both debris size distribution and dynamical behaviour post-impact are critical to estimate the final fraction of the ejecta that will reaccrete, aspects that are not considered in this work. The clump's mass is estimated by gravitational binding without explicit size considerations, and could be considered as an upper limit as some of the fragments can be dragged towards the Sun. However if a consequent fraction of the material re-accretes, higher post-impact $Z_{Fe}$'s than the ones presented in this work are needed to correctly predict Mercury's current $Z_{Fe}$. 

Another important aspect is the fate of the impactor after the collision. When the collision energy is high, part of the colliding material can be unbound and distributed in the neighbouring region. This debris is expected to cool and condense, and can cross Mercury's (or another planet's) orbit and be reaccreted later on. 
If the impactor is larger than Mercury, it likely survives the collision with little mantle stripping. 
Unless the orbit is extremely eccentric which could lead to inward scattering and possibly falling into the Sun's gravitational well, it is expected to survive. 
Therefore it seems that an additional mechanism is required to remove the impactor from Mercury's orbit, and we plan to investigate the dynamical evolution of the debris in future work.
\par

\subsection{Comparison with previous work}

We performed simulations with initial conditions from \cite{Benz2007} and \cite{Asphaug} to compare with their results. 
For run-6 of \cite{Benz2007} with $M_{tar}$=2.25~M$_{\mercury}$, $M_{imp}$=0.225~M$_{\mercury}$, $v_{imp} = 20$ km/s, $\theta=0$, a Mercury analog of 0.92~M$_{\mercury}$ and $Z_{Fe}$=0.61 was inferred, 
while we obtain 0.74~M$_{\mercury}$ and $Z_{Fe}=0.67$. The most successful collision in \cite{Asphaug} ($M_1$= 0.85 R$_{\oplus}$, $M_2$= 4.52~M$_{\mercury}$, $v_{imp}= 3.25 v_{esc}$, $\theta= 34\degree$) results in 1.0~M$_{\mercury}$ and $Z_{Fe}$=0.70, while our simulations infer a Mercury analog of 1.07~M$_{\mercury}$ and $Z_{Fe}$=0.76. For Case-2 we find an agreement at the 5\% level while Case-1 agrees within 20\%. 

Overall, a good agreements is found between our simulations and those of \citet{Asphaug}. There are several possible reasons for the differences between ours and the \citet{Benz2007} simulations. 
First, \citet{Benz2007} used the ANEOS to model the mantle which leads to a higher reference density than the one in the Tillotson EOS used in this work (3.32 g/cm$^3$ for dunite, while 2.7 g/cm$^3$ for basalt). 
This makes the initial bodies $\sim 10\%$ smaller and more condensed, as a result they have a higher gravitational binding energy and are harder to disrupt. 
In order to investigate this effect, we simulate run-6 of \citet{Benz2007} with a modified Tillotson EOS where we substitute the reference density of basalt by the one of dunite. We find a body of 0.81 $M_{\mercury}$ with $Z_{Fe}$ of $\sim 0.76$. Thus the choice of mantle material can explain up to 5-6\% of the mass difference. 
Second, as discussed above, disruptive collisions are highly energetic and a significant part of the material can undergo phase transitions. Accounting for the latent heat thus can change the thermal pressure of the material. 
For the Tillotson EOS (which neglects phase transitions), we obtain a hotter and more pressure-supported material, allowing for more disruption.
Finally, differences in the numerical treatments and analysis methods can also affect the results by a few percent (e.g. \citealt{Genda2015}, \citealt{Genda2018}).

\section{Summary and Conclusions \label{conclusion}}
We investigated formation paths of Mercury including giant impact, hit-and-run, and multiple collisions. 
We presented a large parameter study for these three cases, considering different collision parameters, the sensitivity of the results to impactor's composition, and different initial states (inflated, rotating) of the target.

We find that the two end-members of the range of successful giant impacts are with $b=0.5-0.7$ and $v_{imp}\sim30$~km/s (5-6~$v_{esc}$) and with $b=0.2-0.3$ and $v_{imp}\sim15$~km/s (3-4~$v_{esc}$). 
In the first case, the constraints on both the velocity and the angle are tight, and are not very likely. The latter requires a very small impact angle, but with a more moderate velocity ($v_{imp}\sim$15~km/s). 
In Case-2,  the hit-and-run scenario, the impact velocity is closer to the escape velocity. 
On the other hand, a massive object on a highly eccentric orbit is also somewhat rare, and its origin as well as its fate post-impact still need to be investigated and justified. 
In Case-2, we also find that there is a larger parameter space of possibilities to form Mercury-like planets, but the proto-Mercury has to be 4-5 times more massive than its present mass. For the same impact parameters, disruptive collisions are also more likely than in Case-1 since they are more energetic.
It is difficult to assert which of the cases is more probable. 
Future 
investigations of N-body simulations could put limits on the collision rates and statistics. 
Finally, we also show that Mercury can form via several collisions with less extreme conditions each time. The closer to the most likely statistical values for the impact velocity and angle, the more impacts that are required. If the next collision occurs shortly after the first one, more mantle and cloud-like material mostly composed of silicate but with a small iron fraction from the target) can be stripped. A few collisions happening tightly in time is a more effective scenario for reaching Mercury's high $Z_{Fe}$ and is furthermore also consistent with its surface volatile-rich composition.

Our results can be summarized as follows:
\begin{itemize}
\item A single giant impact or hit-and-run impact require highly tuned impact parameters and velocities to reproduce Mercury's mass and $Z_{Fe}$. There is a somewhat larger parameter space of possibilities in the hit-and-run scenario.
\item The impactor's composition affects the resulting final mass and post-impact iron distribution.
\item The pre-impact state of the target affects the resulting final mass. 
\item A multiple collision scenario escapes the fine-tuning of the geometrical parameters but is constrained by the timing and the volatile-rich composition of Mercury's surface. 
\item Forming Mercury by giant impacts is feasible but difficult.  
\end{itemize}
Mercury's origin remains poorly understood as it combines physical, chemical and dynamical processes that must be coupled. 
The low frequency of metal-rich exoplanets (Mercury-analogs) suggests that forming metal-rich planets requires unique circumstances. 
Therefore, understanding the formation of Mercury can reflect on our understanding of exoplanetary systems.  

\section{Acknowledgements}
We thank our colleagues at the institute and PlanetS for various helpful discussions, and the referee for insightful comments that helped to improve the quality of the paper.
CR acknowledges funding through SNF Grant "Computational Astrophysics" (200020 162930/1). RH acknowledges support from the Swiss National Science Foundation (SNSF) Grant No. 200021\_169054. 
Part of this work has been carried out within the framework of the NCCR PlanetS supported by the Swiss National Science Foundation.
All the simulations were performed on the Piz Daint supercomputer at the Swiss National Supercomputing Centre (CSCS). 

\newpage
\section{Appendix \label{appendix}}

\subsection{Density Profile}
\begin{figure}[h!]
\begin{center}
\includegraphics[width=0.4\textwidth]{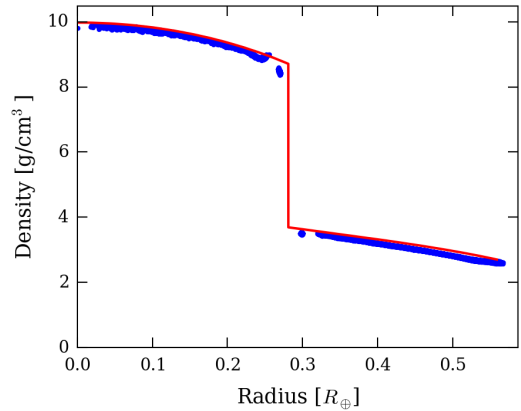}
\caption{ Density profile of a proto-Mercury of $2.25~$M$_{\mercury}$ and with a resolution of 57877 particles after relaxation. The red line shows the solution $\rho(r)$ to the structure equations and the blue dots represent the particle distribution. \label{DensityProfile}}
\end{center}
\end{figure}

Figure \ref{DensityProfile} shows the density profile of proto-Mercury of $2.25~$M$_{\mercury}$. 

\subsection{Scaling Laws \label{scalinglaws}}
In this section we compare our results to scaling laws found by previous studies (e.g., \citealt{Benz1999}, \citealt{Genda2012}, \citealt{LS2012a}). 
A universal scaling law to determine the largest remnant mass after a collision was recently proposed by \cite{LS2012a}: 
\begin{equation}
M_{lr} / M_{tot} = -0.5Q_R / Q^{'*}_{RD} + 1,
\end{equation}
where $Q_R$ is the specific total kinetic energy in the center of mass frame, $Q^{'*}_{RD}$ is the specific energy where half of the target's mass is disrupted (corrected for the interacting mass).

For super-catastrophic collisions, i.e. $M_{lr}/M_{tot}<0.1$, laboratory experiments and simulations indicate that the remnant's mass deviates from the universal scaling law \citep{LS2012a}. 
This regime is better described by the following power law
\begin{equation}
M_{lr}/M_{tot}=\frac{0.1}{1.8^{\eta}}(Q_R/Q^{'*}_{RD})^\eta,
\end{equation}
where $\eta$ is $\sim-1.5$.
Usually one first derives the characteristic energy of the system $Q_{RD}^{'*}$ \citep{Marcus,LS2012a}. 
The calculated $Q_{RD}^{'*}$ following both prescriptions differ from our results by a factor of $\sim$2. 
Due to differences in the numerical treatment the fit to $Q_{RD}^{'*}$ cannot be used, confirming that the critical disruption of a body depends on the details of the numerical method and numerical parameters \citep{Genda2015, Genda2018}. We can estimate $Q_{RD}^{'*}$ from our data for each available subsets of collisions with the same target mass, mass ratio, and impact angle. 
We linearly fit the ratio $M_{lr}/M_{tot}$ as a function of the specific energy $Q_R$, and infer the corresponding specific energy $Q_R$ such that $M_{lr}/M_{tot}=0.5$. Subsets with 
$M_{lr}/M_{tot} <0.3$, and where the available maximum and minimum $M_{lr}/M_{tot}$ differ by less than 10\% are excluded in order to avoid spurious extrapolation. 
For the hit-and-run collisions we substitute the mass ratio $M_{lr}/M_{tot}$ by $M_{2lr}/M_{pM}$ which is the mass ratio between the second largest remnant and the protoplanet's initial mass \citep{LS2012a}.

\begin{figure}[h!]
\includegraphics[width=0.32\textwidth]{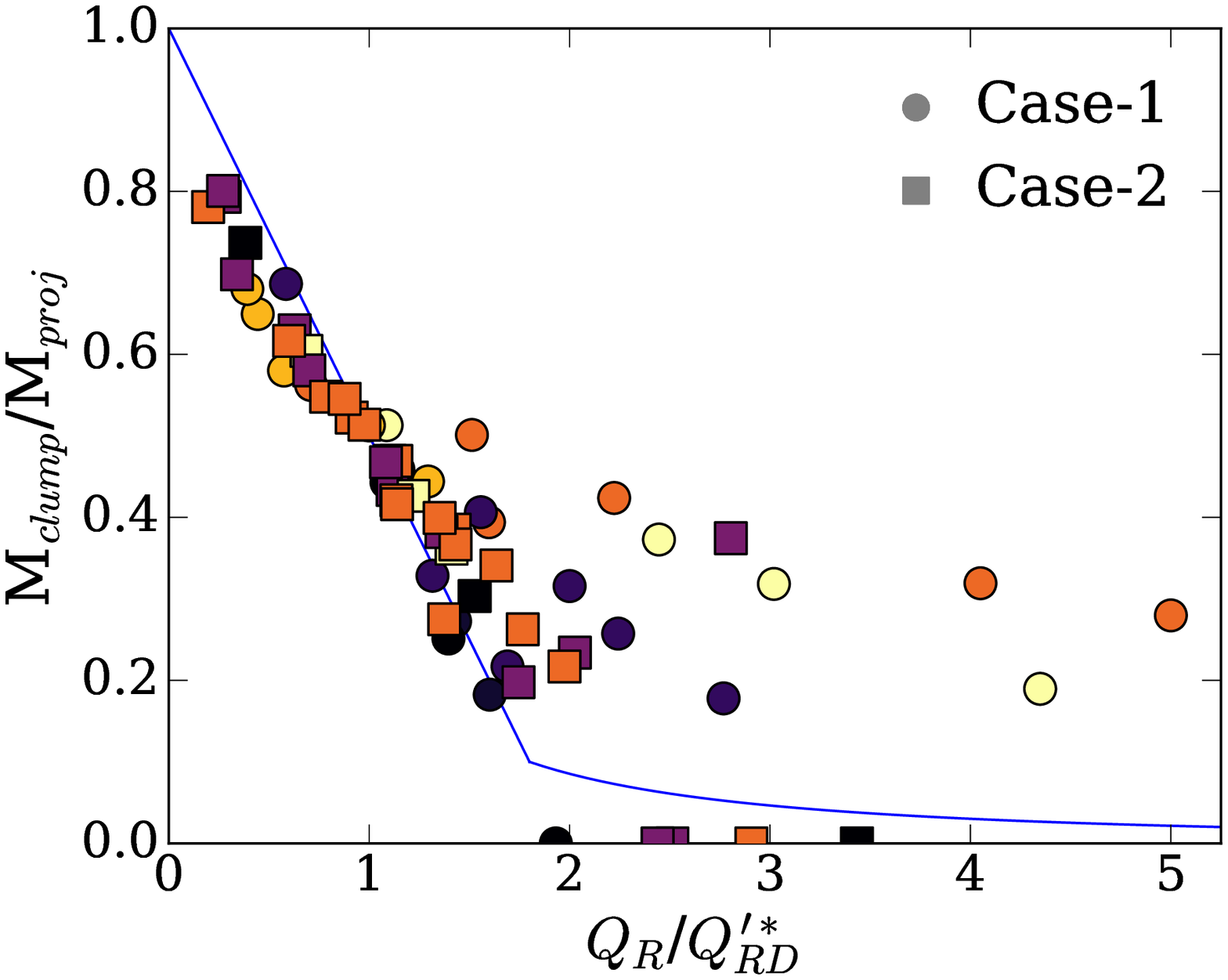}
\includegraphics[width=0.32\textwidth]{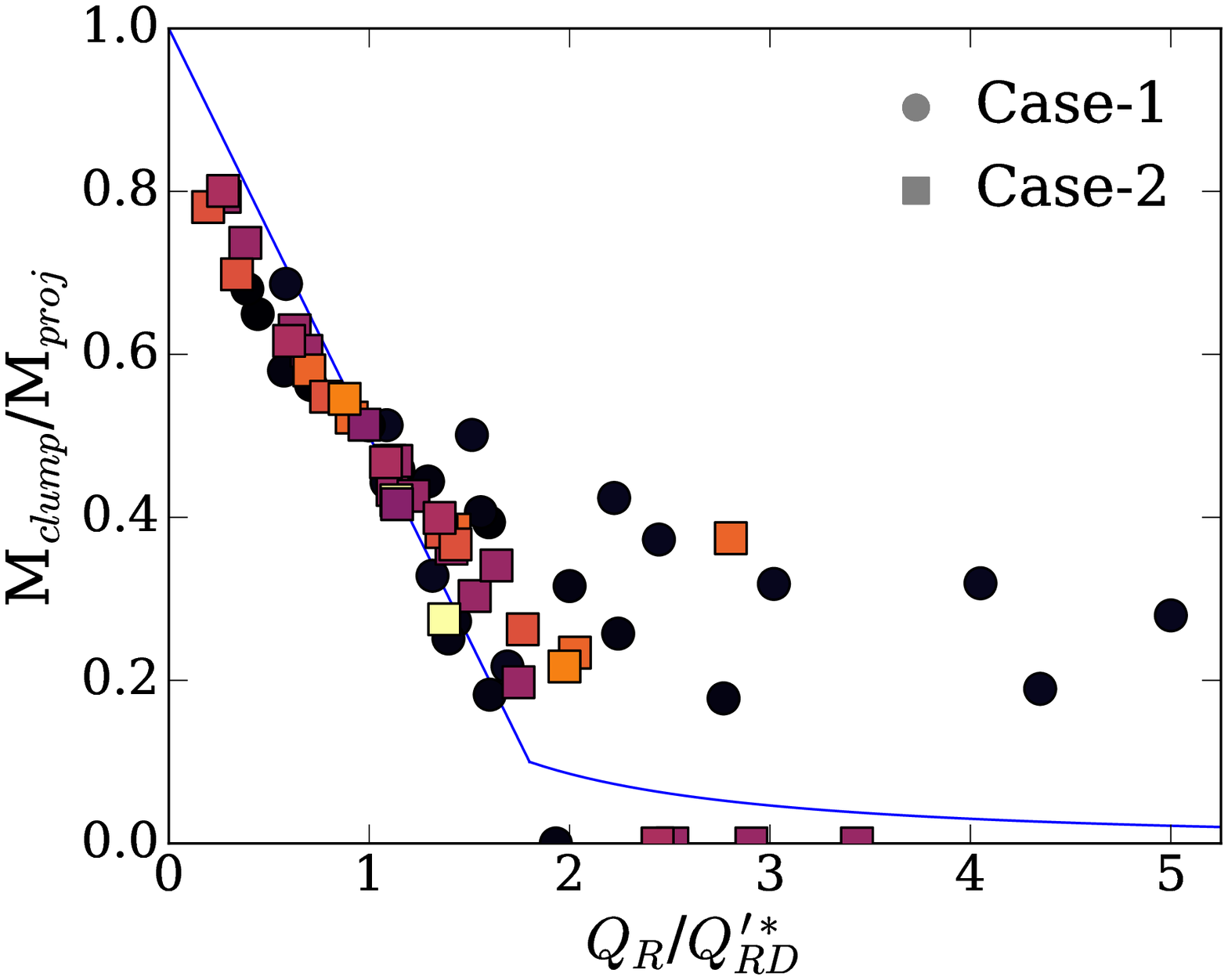}
\includegraphics[width=0.32\textwidth]{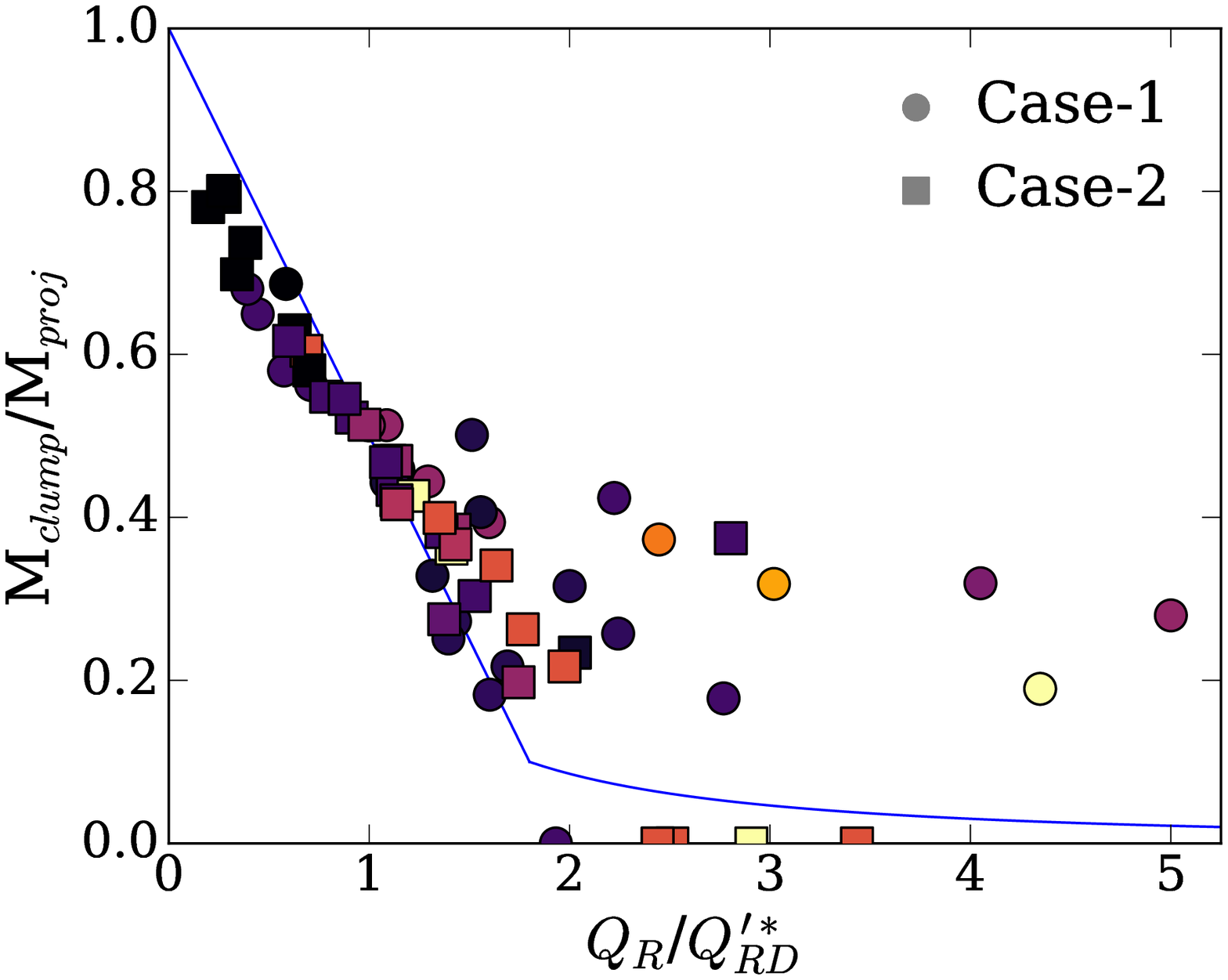}

\includegraphics[width=0.32\textwidth]{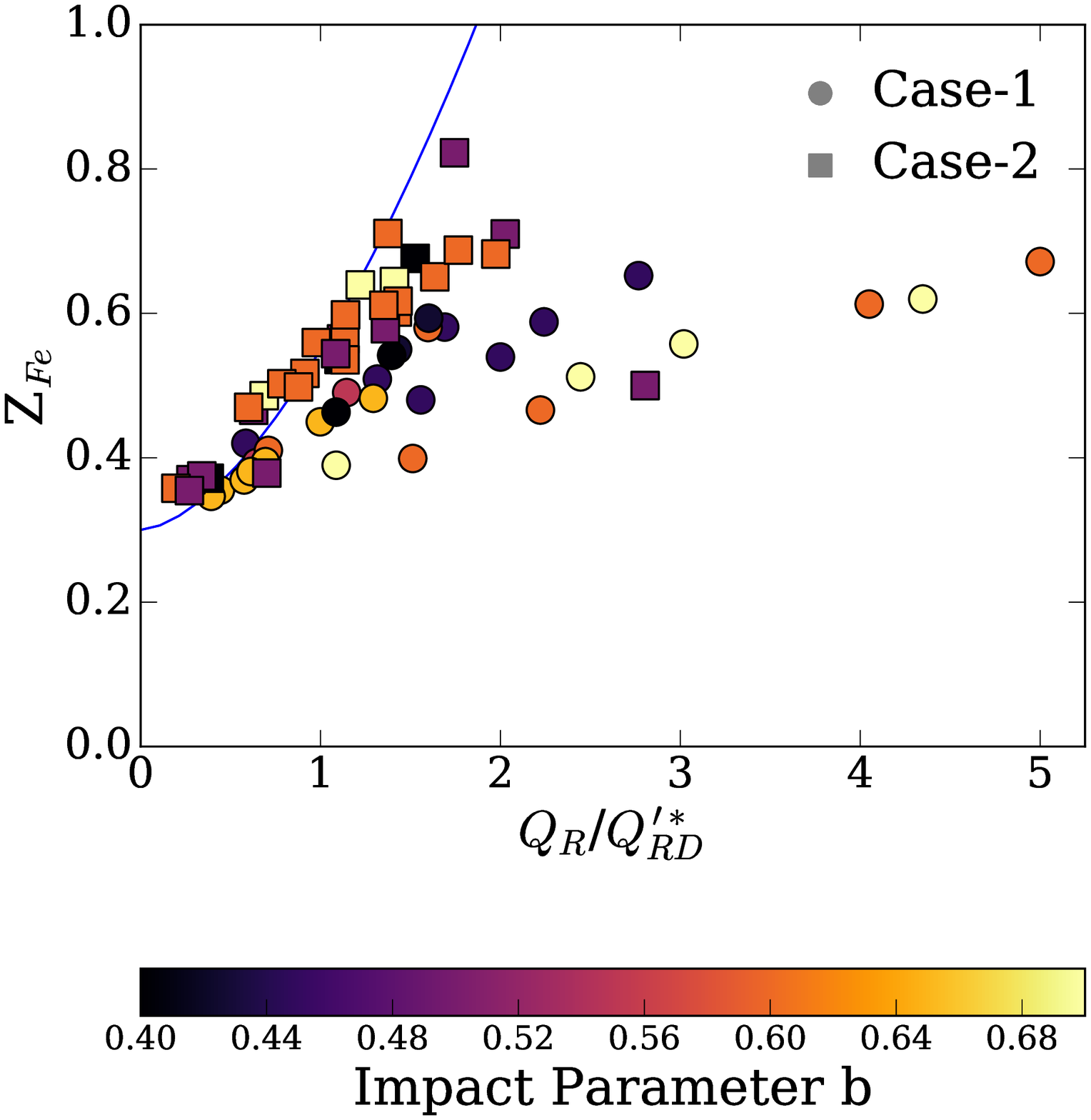}
\includegraphics[width=0.32\textwidth]{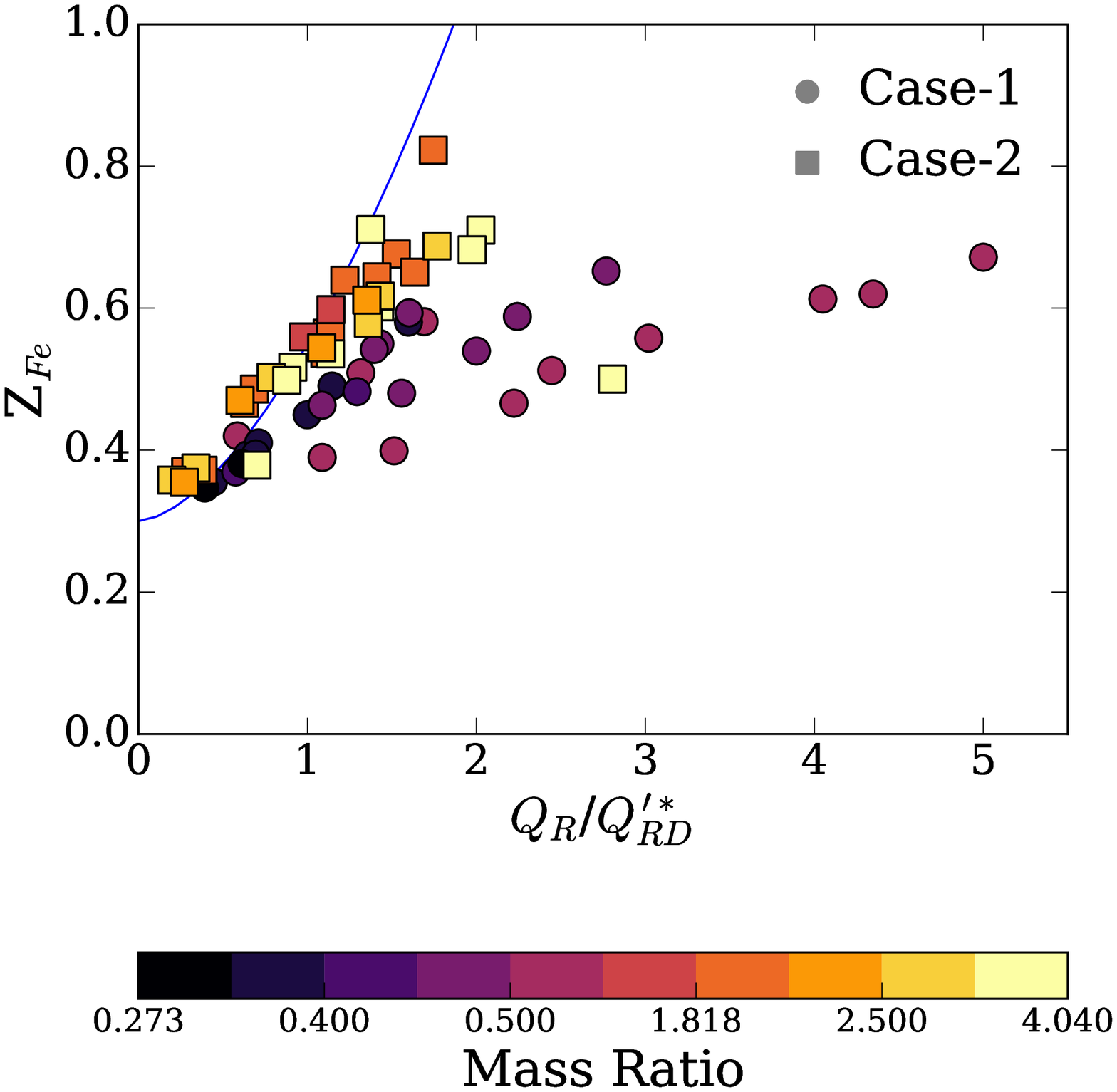}
\includegraphics[width=0.32\textwidth]{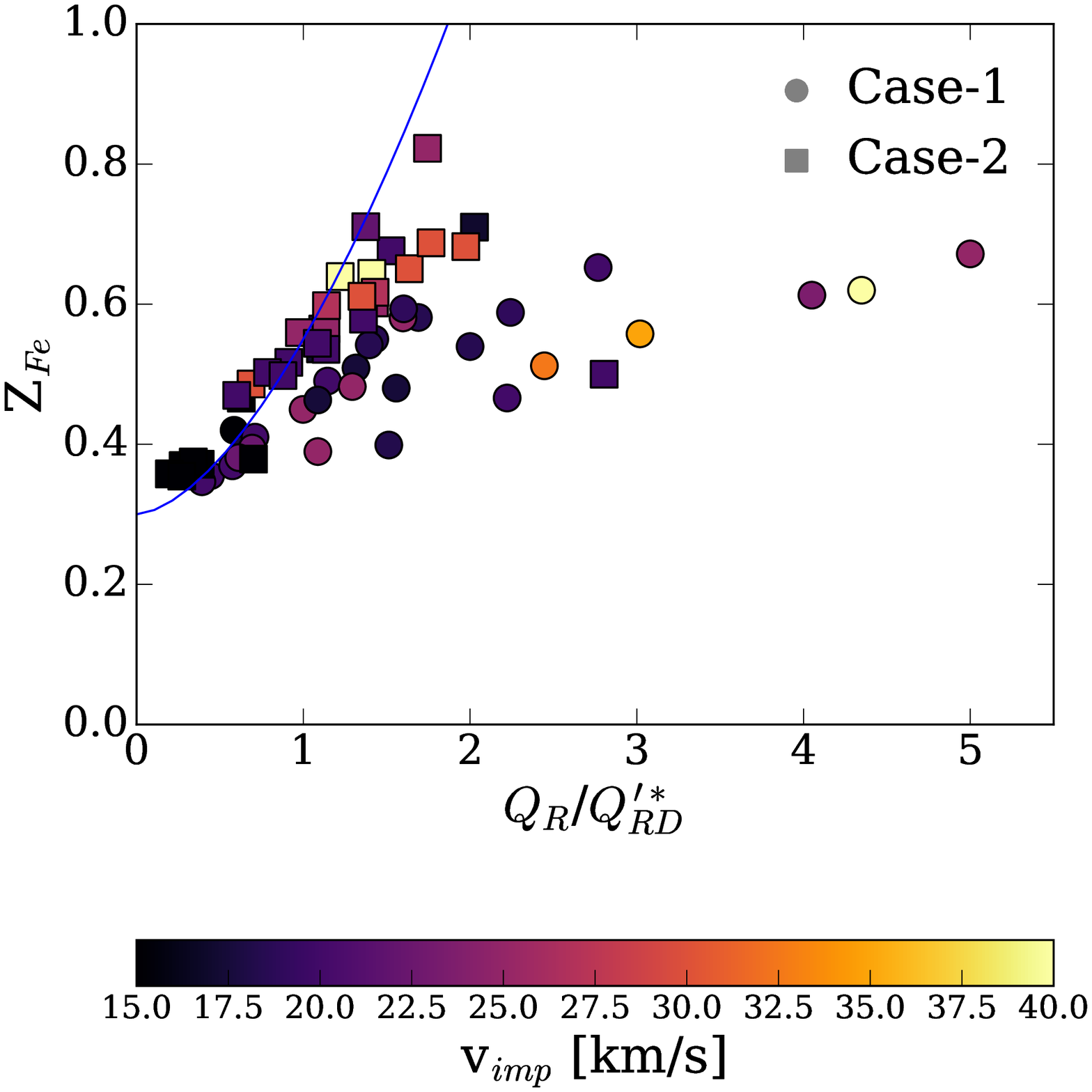}

\caption{Scaling laws for Case-1 (circles) and Case-2 (squares). Top: $M_{clump}/M_{tot}$ vs.~the characteristic energy ratio ($Q_R/Q_{RD}^{'*}$). The blue line corresponds to the universal scaling law with the super-catastrophic regime \citep{LS2012a}. Bottom: $Z_{Fe}$ in the fragment vs.~the energy ratio. The blue line presents the scaling law of \cite{Marcus}, scaled down for our initial iron mass fraction.
The color maps indicate the collision parameters: (left) impact parameter (middle) projectile-to-target mass ratio (right) impact velocity in km/s. \label{scaling}}
\end{figure}

For the inferred $Z_{Fe}$, \cite{Marcus} proposed the following power law:
\begin{equation}
M_{Fe}/M_{lr}=0.3+0.25(Q_R/Q^{'*}_{RD}-1)^{1.65}, 
\end{equation}
where the numerical factor can be scaled down to fit our initial core fraction of 0.3 instead of 0.33 \citep{Carter2018}. 
The upper panels of Figure \ref{scaling} show $M_{clump}/M_{tot}$ and $M_{2lr}/M_{pM}$, as a function of the specific energy normalized by the characteristic specific energy $Q_R/Q_{RD}^{'*}$ for Case-1 and Case-2, respectively. 
The blue line represents the universal scaling law of \citet{LS2012a}. The lower panels of Figure \ref{scaling} show $Z_{Fe}$  as a function of $Q_R/Q_{RD}^{'*}$ for Case-1 (circles) and Case-2 (squares).

 For low energies, we indeed find that the mass ratio is linearly dependent on the energy ratio. For $Q_R/Q^{'*}_{RD}>$1.7, we find a significant scatter, especially for Case-1. 
 This might be due to the transition to the super-catastrophic regime. We also note that our results for Case-1 have a more moderate slope because we do not probe the low range of energies, which would correct the extrapolated $Q^{'*}_{RD}$. We also note some deviations from the universal slope between subsets depending on the impact angle, specially for more oblique collisions in agreement with \cite{LS2012a}. 
We also find good agreement with the power law of \cite{Marcus} where the mass ratio is described by the linear relation to $Q_R/Q^{'*}_{RD}$. This is especially noticeable 
for Case-2 that follow both scaling laws except for a few outliners. 
For higher energies, we find a similar trend in the $Z_{Fe}$ scatter as in the mass ratio scatter. While the precise value of $Q^{'*}_{RD}$ depends on several numerical parameters, we find a general agreement with the proposed scaling laws once $Q^{'*}_{RD}$ is obtained from simulation data.

\newpage

\subsection{Performed Simulations}
\subsubsection{Case-1}
Table \ref{TableCase1} shows the simulation initial conditions and outcomes for all the Case-1 collisions.
\begin{longtable}{l|l|l|l|l|l || l|l}
   $M_{p\mercury}$ [M$_{\mercury}$] & $N_{p\mercury}$ & $M_{imp}$ [M$_{\mercury}$]& $N_{imp}$ & $b$ & $v_{imp}$ [km/s] & $M_{lr}$ [M$_{\mercury}$] & $Z_{Fe}$\\ \hline \hline
    2.25    & 57877 & 0.675 & 17473 & 0.4   & 15    & 1.72  & 0.39  \\
    2.25    & 57877 & 0.675 & 17473 & 0.4   & 20    & 1.35  & 0.49  \\
    2.25    & 57877 & 0.675 & 17473 & 0.48  & 30    & 1.35  & 0.49  \\ 
    2.25    & 57877 & 0.675 & 17473 & 0.5   & 20    & 1.35  & 0.49  \\
    2.25    & 57877 & 0.675 & 17473 & 0.5   & 30    & 1.35  & 0.49  \\ 
    2.25    & 57877 & 0.675 & 17473 & 0.6   & 20    & 1.35  & 0.49  \\ 
    2.25    & 57877 & 0.675 & 17473 & 0.6   & 25    & 1.35  & 0.49  \\ 
    2.25    & 57877 & 0.675 & 17473 & 0.6   & 30    & 1.35  & 0.49
    \\ \hline 
    2.25    & 57877 & 0.9 & 20953 & 0.6   & 30    & 1.35  & 0.49  \\ 
    2.25    & 57877 & 0.9 & 20953 & 0.6   & 30    & 1.35  & 0.49  \\ 
    2.25    & 57877 & 0.9 & 20953 & 0.6   & 30    & 1.35  & 0.49  \\ 
    2.25    & 57877 & 0.9 & 20953 & 0.6   & 30    & 1.35  & 0.49  \\ 
    2.25    & 57877 & 0.9 & 20953 & 0.6   & 30    & 1.35  & 0.49
    \\ \hline
    2.25    & 57877 & 1.125 & 21205 & 0.1   & 15    & 0.86  & 0.5   \\ 
    2.25    & 57877 & 1.125 & 21205 & 0.2   & 10    & 2.32  &  0.42     \\ 
    2.25    & 57877 & 1.125 & 21205 & 0.2   & 15    & 1.11  & 0.51  \\ 
    2.25    & 57877 & 1.125 & 21205 & 0.2   & 17    & 0.73  & 0.58  \\ 
    2.25    & 57877 & 1.125 & 21205 & 0.25  & 20    & 0.60  & 0.68  \\ 
    2.25    & 57877 & 1.125 & 21205 & 0.3   & 15    & 1.07  & 0.55  \\ 
    2.25    & 57877 & 1.125 & 21205 & 0.3   & 20    & 0.81  & 0.64  \\ 
    2.25    & 57877 & 1.125 & 21205 & 0.4   & 20    & 1.15  & 0.55  \\ 
    2.25    & 57877 & 1.125 & 21205 & 0.5   & 16.5  & 1.70  &  0.40    \\ 
    2.25    & 57877 & 1.125 & 21205 & 0.5   & 20    & 1.43  & 0.47  \\ 
    2.25    & 57877 & 1.125 & 21205 & 0.5   & 27    & 1.08  & 0.61  \\ 
    2.25    & 57877 & 1.125 & 21205 & 0.5   & 30    & 0.95  & 0.67  \\ 
    2.25    & 57877 & 1.125 & 21205 & 0.51  & 30    & 0.96  & 0.66  \\ 
    2.25    & 57877 & 1.125 & 21205 & 0.6   & 15    & 1.32  & 0.51  \\ 
    2.25    & 57877 & 1.125 & 21205 & 0.7   & 30    & 1.74  & 0.39  \\ 
    2.25    & 57877 & 1.125 & 21205 & 0.7   & 45    & 1.26  & 0.51  \\ 
    2.25    & 57877 & 1.125 & 21205 & 0.7   & 50    & 1.08  & 0.56  \\ 
    2.25    & 57877 & 1.125 & 21205 & 0.7   & 60    & 0.64  & 0.62
    \\ \hline
    2.475   & 58645 & 0.675 & 17473 & 0.45  & 30    & 1.18  & 0.62  \\ 
    2.475   & 58645 & 0.675 & 17473 & 0.6   & 20    & 2.15  & 0.35  \\ 
    2.475   & 58645 & 0.675 & 17473 & 0.6   & 25    & 1.95  & 0.38
    \\ \hline
    2.475   & 58645 & 0.9 & 20953 & 0.4     & 25    & 1.13  & 0.63  \\ 
    2.475   & 58645 & 0.9 & 20953 & 0.4     & 30    & 0.87  & 0.76  \\ 
    2.475   & 58645 & 0.9 & 20953 & 0.42    & 28    & 1.04  & 0.68  \\ 
    2.475   & 58645 & 0.9 & 20953 & 0.42    & 29    & 0.98  & 0.71  \\ 
    2.475   & 58645 & 0.9 & 20953 & 0.42    & 30    & 0.93  & 0.73  \\ 
    2.475   & 58645 & 0.9 & 20953 & 0.43    & 30    & 0.95  & 0.72  \\ 
    2.475   & 58645 & 0.9 & 20953 & 0.435   & 30    & 0.97  & 0.71  \\ 
    2.475   & 58645 & 0.9 & 20953 & 0.437   & 30    & 0.97  & 0.71
    \\ \hline
    2.475   & 58645 & 1.125 & 21205 & 0.2   & 17    & 1.00  & 0.56  \\ 
    2.475   & 58645 & 1.125 & 21205 & 0.43  & 27    & 1.02  & 0.68  \\ 
    2.475   & 58645 & 1.125 & 21205 & 0.45  & 27    & 1.10  & 0.65  \\ 
    2.475   & 58645 & 1.125 & 21205 & 0.45  & 29    & 0.96  & 0.71  \\ 
    2.475   & 58645 & 1.125 & 21205 & 0.45  & 30    & 0.92  & 0.73  \\ 
    2.475   & 58645 & 1.125 & 21205 & 0.47  & 30    & 0.98  & 0.69  \\ 
    2.475   & 58645 & 1.125 & 21205 & 0.5   & 30    & 1.13  & 0.64
    \\ \hline
    2.7     & 68905 & 1.125 & 21205 & 0.1   & 15    & 1.70  & 0.46  \\ 
    2.7     & 68905 & 1.125 & 21205 & 0.1   & 17    & 0.96  & 0.54  \\ 
    2.7     & 68905 & 1.125 & 21205 & 0.1   & 20    & 0  & 0  \\ 
    2.7     & 68905 & 1.125 & 21205 & 0.15  & 17    & 1.04  & 0.55  \\ 
    2.7     & 68905 & 1.125 & 21205 & 0.15  & 18    & 0.70  & 0.59  \\ 
    2.7     & 68905 & 1.125 & 21205 & 0.2   & 15    & 1.56  & 0.48  \\ 
    2.7     & 68905 & 1.125 & 21205 & 0.2   & 17    & 1.21  & 0.54  \\ 
    2.7     & 68905 & 1.125 & 21205 & 0.2   & 18    & 0.99  & 0.59  \\ 
    2.7     & 68905 & 1.125 & 21205 & 0.2   & 20    & 0.68  & 0.65  \\ 
    2.7     & 68905 & 1.125 & 21205 & 0.3   & 20    & 1.18  & 0.58  \\ 
    2.7     & 68905 & 1.125 & 21205 & 0.3   & 23    & 0.90  & 0.67  \\ 
    2.7     & 68905 & 1.125 & 21205 & 0.3   & 25    & 0.68  & 0.74  \\ 
    2.7     & 68905 & 1.125 & 21205 & 0.43  & 30    & 1.01  & 0.73  \\ 
    2.7     & 68905 & 1.125 & 21205 & 0.45  & 30    & 1.09  & 0.69  \\ 
    2.7     & 68905 & 1.125 & 21205 & 0.5   & 30    & 1.32  & 0.61  \\ 
    2.7     & 68905 & 1.125 & 21205 & 0.7   & 30    & 1.41  & 0.53  
    \\ \hline \hline
    \caption{The initial conditions (initial proto-Mercury's and impactor's masses, $M_{p\mercury}$, $M_{imp}$ and resolutions $N_{p\mercury}$, $N_{imp}$, the impact parameter $b$, the impact velocity $v_{imp}$) and outcomes (mass of the largest fragment $M_{lr}$ and its iron mass fraction $Z_{Fe}$) for Case-1 collisions. }
    \label{TableCase1}
\end{longtable}

\subsubsection{Case-2}
Table \ref{TableCase2} shows the simulation initial conditions and outcomes for all the Case-2 collisions.
\begin{longtable}{l|l|l|l|l|l || l|l}
   $M_{p\mercury}$ [M$_{\mercury}$] & $N_{p\mercury}$ & $M_{imp}$ [M$_{\mercury}$]& $N_{imp}$ & $b$ & $v_{imp}$ [km/s] & $M_{2lr}$ [M$_{\mercury}$] & $Z_{Fe}$\\ \hline \hline
   2.475   & 58645 & 4.53 & 108901 & 0.2   & 10    & 5.83  & 0.35 \\ 
   2.475   & 58645 & 4.53 & 108901 & 0.3   & 20    & 0.54  & 0.70 \\ 
   2.475   & 58645 & 4.53 & 108901 & 0.4   & 10    & 1.83  & 0.37 \\
   2.475   & 58645 & 4.53 & 108901 & 0.4   & 17    & 1.16  & 0.54 \\
   2.475   & 58645 & 4.53 & 108901 & 0.4   & 20    & 0.75  & 0.68 \\
   2.475   & 58645 & 4.53 & 108901 & 0.4   & 30    & 0  & 0 \\ 
   2.475   & 58645 & 4.53 & 108901 & 0.5   & 10    & 1.97  & 0.37 \\ 
   2.475   & 58645 & 4.53 & 108901 & 0.5   & 15    & 0.54  & 0.47 \\ 
   2.475   & 58645 & 4.53 & 108901 & 0.5   & 20    & 1.08  & 0.56 \\ 
   2.475   & 58645 & 4.53 & 108901 & 0.5   & 25    & 0.49  & 0.82 \\ 
   2.475   & 58645 & 4.53 & 108901 & 0.6   & 25    & 1.16  & 0.56 \\ 
   2.475   & 58645 & 4.53 & 108901 & 0.6   & 30    & 0.85  & 0.65 \\ 
   2.475   & 58645 & 4.53 & 108901 & 0.6   & 40    & 0  & 0 \\ 
   2.475   & 58645 & 4.53 & 108901 & 0.7   & 30    & 1.50  & 0.49 \\
   2.475   & 58645 & 4.53 & 108901 & 0.7   & 40    & 1.06  & 0.64\\ \hline 
   2.475   & 58645 & 6.75 & 167329 & 0.4   & 10    & 0  & 0 \\ 
   2.475   & 58645 & 6.75 & 167329 & 0.4   & 20    & 0  & 0 \\ 
   2.475   & 58645 & 6.75 & 167329 & 0.5   & 10    & 1.44  & 0.38 \\
   2.475   & 58645 & 6.75 & 167329 & 0.5   & 17    & 0.58  & 0.71 \\
   2.475   & 58645 & 6.75 & 167329 & 0.5   & 20    & 0.93  & 0.50 \\
   2.475   & 58645 & 6.75 & 167329 & 0.6   & 20    & 1.30  & 0.52 \\
   2.475   & 58645 & 6.75 & 167329 & 0.6   & 25    & 0.96  & 0.60 \\
   2.475   & 58645 & 6.75 & 167329 & 0.7   & 30    & 1.35  & 0.47 \\ \hline
   2.475   & 58645 & 10 & 238777 & 0.4   & 10    & 0  & 0 \\ 
   2.475   & 58645 & 10 & 238777 & 0.5   & 10    & 0  & 0 \\ 
   2.475   & 58645 & 10 & 238777 & 0.5   & 20    & 0  & 0 \\ 
   2.475   & 58645 & 10 & 238777 & 0.6   & 20    & 1.04  & 0.54 \\ 
   2.475   & 58645 & 10 & 238777 & 0.6   & 22    & 0.68  & 0.71 \\ 
   2.475   & 58645 & 10 & 238777 & 0.6   & 30    & 0  & 0 \\ 
   2.475   & 58645 & 10 & 238777 & 0.7   & 30    & 1.16  & 0.59 \\ \hline
   2.7     & 68905 &  4.53 & 10890 & 0.4 	& 30	& 	0 & 0 \\
   2.7     & 68905 &  4.53 & 10890 & 0.5 	& 30	& 	0 & 0 \\
   2.7     & 68905 &  4.53 & 10890 & 0.6 	& 25	& 	1.39 & 0.56  \\
   2.7     & 68905 &  4.53 & 10890 & 0.6 	& 27	& 	1.13 & 0.60 \\
   2.7     & 68905 &  4.53 & 10890 & 0.7 	& 40	& 	1.33 & 0.60 \\ \hline
   2.7     & 68905 & 6.75 & 167329 & 0.5   & 10    & 1.89  & 0.37  \\ 
   2.7     & 68905 & 6.75 & 167329 & 0.5   & 20    & 1.03  & 0.58  \\ 
   2.7     & 68905 & 6.75 & 167329 & 0.6   & 10    & 2.11  & 0.36  \\ 
   2.7     & 68905 & 6.75 & 167329 & 0.6   & 20    & 1.48  & 0.50  \\ 
   2.7     & 68905 & 6.75 & 167329 & 0.6   & 27    & 0.99  & 0.62  \\ 
   2.7     & 68905 & 6.75 & 167329 & 0.6   & 30    & 0.71  & 0.69  \\ 
   2.7     & 68905 & 6.75 & 167329 & 0.7   & 30    & 1.54  & 0.51  \\ \hline 
   2.7     & 68905 & 10 & 238777 & 0.6   & 20    & 1.32  & 0.52  \\ 
   2.7     & 68905 & 10 & 238777 & 0.6   & 30    & 0  & 0  \\
   2.7     & 68905 & 10 & 238777 & 0.7   & 30    & 1.37  & 0.56  \\ \hline
   3.375   & 85573 & 6.75 & 167329 & 0.5   & 10    & 2.7   & 0.35 \\
   3.375   & 85573 & 6.75 & 167329 & 0.5   & 20    & 1.58  & 0.54 \\
   3.375   & 85573 & 6.75 & 167329 & 0.5   & 30    & 0     & 0 \\
   3.375   & 85573 & 6.75 & 167329 & 0.6   & 20    & 2.09   & 0.47 \\
   3.375   & 85573 & 6.75 & 167329 & 0.6   & 30    & 1.35   & 0.61 \\
   3.375   & 85573 & 6.75 & 167329 & 0.7   & 30    & 2.17   & 0.46 \\ \hline
   3.375   & 85573 & 10 & 238777 & 0.5   & 10    & 0   & 0 \\
   3.375   & 85573 & 10 & 238777 & 0.5   & 20    & 0   & 0 \\
   3.375   & 85573 & 10 & 238777 & 0.5   & 30    & 0   & 0 \\
   3.375   & 85573 & 10 & 238777 & 0.6   & 20    & 1.84   & 0.50 \\
   3.375   & 85573 & 10 & 238777 & 0.6   & 30    & 0.73   & 0.68 \\
   3.375   & 85573 & 10 & 238777 & 0.7   & 30    & 1.96   & 0.50 \\
    \caption{The initial conditions (initial proto-Mercury's and impactor's masses, $M_{p\mercury}$, $M_{imp}$ and resolutions $N_{p\mercury}$, $N_{imp}$, the impact parameter $b$, the impact velocity $v_{imp}$) and outcomes (mass of the second largest fragment $M_{2lr}$ and its iron mass fraction $Z_{Fe}$) for Case-2 collisions.}
    \label{TableCase2}
   
\end{longtable}

\subsubsection{Case-3}
Table \ref{TableCase3} shows the simulation initial conditions and outcomes for all the Case-3 collisions.
\begin{longtable}{l|l|l|l|l|l || l|l}
   $M_{p\mercury}$ [M$_{\mercury}$] & $N_{p\mercury}$ & $M_{imp}$ [M$_{\mercury}$] & $N_{imp}$ & $b$ & $v_{imp}$ [km/s] & $M_{lr}$ [M$_{\mercury}$] & $Z_{Fe}$\\ \hline \hline
   2.25   & 57877 & 0.45 & 11653 & 0.7   & 13.6   & 2.22  & 0.31 \\ 
   2.22   & 58273 & 0.45 & 11653 & 0.7   & 13.5   & 2.14  & 0.32 \\ 
   2.14   & 77185 & 0.45 & 11653 & 0.7   & 13.4   & 2.08  & 0.32 \\ 
   2.08   & 55081 & 0.45 & 11653 & 0.7   & 13.4   & 2.03  & 0.33 \\
   2.03   & 52585 & 0.45 & 11653 & 0.7   & 13.2   & 1.97  & 0.34 \\
   1.97   & 49369 & 0.45 & 11653 & 0.7   & 13.1   & 1.92  & 0.35 \\
   1.92   & 49369 & 0.45 & 11653 & 0.7   & 13.1   & 1.86  & 0.36 \\
   1.86   & 45037 & 0.45 & 11653 & 0.7   & 12.9   & 1.81  & 0.37 \\
   1.81   & 45481 & 0.45 & 11653 & 0.7   & 12.8   & 1.75  & 0.39 \\
   1.75   & 40285 & 0.45 & 11653 & 0.7   & 12.7   & 1.69  & 0.40 \\
   1.69   & 40705 & 0.45 & 11653 & 0.7   & 12.6   & 1.64 & 0.41 \\
   1.64   & 41149 & 0.45 & 11653 & 0.7   & 13.4   & 1.58  & 0.43 \\
   1.58   & 39865 & 0.45 & 11653 & 0.7   & 13.4   & 1.53  & 0.44 \\
   1.53   & 40705  & 0.45 & 11653 & 0.7   & 13.4   & 1.48  & 0.46 \\
   1.48   & 35233  & 0.45 & 11653 & 0.7   & 13.4   & 1.33  & 0.47 \\
   1.43   & 36397 & 0.45 & 11653 & 0.7   & 13.4   & 1.38  & 0.49 \\
   1.38   & 32929  & 0.45 & 11653 & 0.7 & 11.9	& 1.33 & 0.51\\
\\ \hline
2.25 & 57877	 & 	0.45 & 11653 &	 	0.7 & 18.1 &		2.15 & 	0.31 \\
2.15 & 50281 & 	0.45 & 11653 & 	0.7 & 17.9 & 		2.05 & 	0.33 \\
2.05	 & 51721 & 	0.45 & 11653 &	 	0.7 &	17.7 & 	1.94	& 	0.35 \\
1.94 & 49369 &		0.45 & 11653 &	 	0.7	& 17.4	 &	1.84	& 	0.37 \\
1.84	 & 45481 & 	0.45 & 11653 & 	0.7 &	17.1	 &	1.74	& 		0.39 \\
1.74 & 40285  &	 0.45 & 11653 & 	0.7 &	16.9	 &	1.63	& 		0.41 \\
1.63 & 42121 &	 	0.45 & 11653 & 	0.7 &	16.6		&	1.53	& 		0.44 \\
1.53 & 40705 &	 	0.45 & 11653 &		0.7 &	16.3		&	1.44	& 		0.47 \\
1.44 & 35605 & 	0.45 & 11653 &		0.7 &	16.1		&	1.33 & 		0.52 \\
1.33 & 32929 & 	0.45 & 11653 &		0.7 &	15.7		&	1.24	& 		0.54 \\
1.24	 & 32929 &	0.45	 & 11653 & 	0.7 &	15.5		&	1.16	& 		0.58 \\
1.16	 & 28069 &	0.45 & 11653 &	 	0.7 &	15.2		&	1.08	& 		0.62 \\
1.08 & 29089 & 	0.45	 & 11653 & 	0.7 &	15.1		&	1.01	& 		0.67 \\ \hline
2.25 & 57877 &		0.675 &  17473	&	0.7 &	13.8 & 2.19 &	0.31 \\
2.19	& 54637 &		0.675	& 17473	&	0.7 &	13.7	 & 2.11 &	0.32 \\
2.11	& 55081 &		0.675	& 17473 &	0.7 &	13.6	 &2.03 & 	0.34 \\
2.03	& 52585 &		0.675	& 17473 & 0.7 &	13.5	 & 1.86 &	0.37 \\
1.86	& 45481 &		0.675	& 17473	&	0.7 &	13.2	 & 1.78 &	0.38 \\
1.78	& 40285 &		0.675	& 17473	&	0.7 &	13.0	 & 1.70 &	0.40 \\
1.70	& 40705 &		0.675	& 17473	&	0.7 &	12.9	 & 1.62 &	0.42 \\
1.62	& 39865 &		0.675	& 17473	&	0.7 &	12.7	 & 1.54 &	0.44 \\
1.54	& 40705 &		0.675	& 17473	&	0.7 &	12.6	 & 1.46 &	0.47 \\
1.46	& 35605 &		0.675	& 17473	&	0.7 &	12.4	 & 1.39 &	0.49 \\
1.39	& 32929 &		0.675	& 17473	&	0.7 &	12.3	 & 1.31 &	0.52 \\
1.31	& 32929 &		0.675	& 17473	&	0.7 &	12.2	 & 1.24 &	0.55 \\
1.24	& 32533 &		0.675	& 17473	&	0.7 &	12.0	 & 1.16 &	0.59 \\
1.16	& 28393 &		0.675	& 17473	&	0.7 &	11.8	 & 1.09 &	0.62 \\
1.09	& 29089 &		0.675	& 17473	&	0.7 &	11.8	 & 1.02 &	0.66 \\ \hline
2.25 & 57877 &		0.675 &  17473	&	0.7 &	18.4 & 2.11 & 	0.57 \\
2.11	& 55081 &		0.675 &  17473	&	0.7 &	18.2 & 1.96 &	0.34	\\
1.96 & 49369 &		0.675 &  17473	&	0.7 &	17.9 & 1.81 &	0.37	\\
1.81	& 45481 &		0.675 &  17473	&0.7 &	17.5 & 1.67 &	0.40	\\
1.67	& 42337 &		0.675 &  17473	&0.7 &	17.1	& 1.53 &		0.44	\\
1.53	& 40705 &		0.675 &  17473	&0.7 &	16.8	& 1.40 &		0.48\\
1.40	& 36397 &		0.675 &  17473	&0.7 &	16.5	& 1.28 &		0.53	\\
1.28	& 31789 &		0.675 &  17473	&0.7 &	16.2	& 1.16 &		0.58	\\
1.16& 57877 &		0.675 &  17473	&0.7 &	15.9	 & 1.06 &	0.64 \\ \hline
2.25 & 57877 &		1.125 &	21205 &	0.8 &	19.0 & 2.17 &	0.31 \\
2.17 & 54637 &		1.125 &	21205 &	0.8 &	18.9	& 2.07 &	0.33 \\
2.07 & 55081 &		1.125 &	21205 &	0.8 &	18.8	& 1.97 &	0.34 \\
1.97	& 49369 &		1.125 &	21205 &	0.8 &	18.5	& 1.87 &	0.36 \\
1.87	& 45037 &		1.125 &	21205 &	0.8 &	18.3	& 1.77 &	0.38 \\
1.77	& 40285 &		1.125 &	21205 &	0.8 &	18.1	& 1.68 &	0.40 \\
1.68	& 44593 &		1.125 &	21205 &	0.8 &	18.0	 & 1.59 &	0.43 \\
1.59	& 39865 &		1.125 &	21205 &	0.8 &	17.8	& 1.50 &	0.45 \\
1.50	& 40705 &		1.125 &	21205 &	0.8 &	17.6	& 1.41 &	0.48 \\
1.41	& 35605 &		1.125 &	21205 &	0.8 &	17.4 & 1.33 &	0.52 \\
1.33	& 32929 &		1.125 &	21205 &	0.8 &	17.2	& 1.24 &	0.55 \\
1.24	& 32533 &		1.125 &	21205 &	0.8 &	16.9	& 1.16 &	0.58 \\
1.16	& 27769 &		1.125 &	21205 &	0.8 &	16.7	& 1.10 &	0.62 \\ \hline
2.25 & 57877 &		1.125 &	21205 &	0.8 &	14.2	& 2.19 &	0.31 \\
2.19	& 58273 &		1.125 &	21205 &	0.8 &	14.2	& 2.12 &	0.32 \\
2.12	& 50281 &		1.125 &	21205 &	0.8 &	14.1 & 2.06 &	0.33 \\
2.06	& 55081 &		1.125 &	21205 &	0.8 &	14.0	& 2.00 & 	0.34 \\
2.00	& 52585 &		1.125 &	21205 &	0.8 &	14.0	& 1.96 &	0.34 \\
1.96	& 49369 &		1.125 &	21205 &	0.8 &	13.9	& 1.90  &	0.35 \\
1.90	& 49369 &		1.125 &	21205 &	0.8 &	13.8	& 1.83 &	0.37 \\
1.83	& 42337 &		1.125 &	21205 &	0.8 &	13.7	& 1.77 &	0.38 \\
1.77	& 40285 &		1.125 &	21205 &	0.8 &	13.5	& 1.71 &	0.40 \\
1.71	& 40705 &		1.125 &	21205 &	0.8 &	13.5	& 1.64 &	0.41 \\
1.64	& 44593 &		1.125 &	21205 &	0.8 &	13.4	& 1.58 &	0.43 \\
1.58	& 39865 &		1.125 &	21205 &	0.8 &	13.3	& 1.53 &	0.44 \\ \hline
%
2.25 & 57877 &		1.125 &	21205 &	0.7 &	19.0	& 2.04 &	0.33 \\	
2.04	& 51721 &		1.125 &	21205 &	0.7 &	18.7	& 1.82 &	0.37 \\
1.82	& 45481 &		1.125 &	21205 &	0.7 &	18.1	& 1.58 &	0.43 \\
1.58	& 39865 &		1.125 &	21205 &	0.7 &	17.8	& 1.38 &	0.49 \\
1.38	& 35605 &		1.125 &	21205 &	0.7 &	17.3	& 1.20 &	0.56 \\
1.20	& 32533 &		1.125 &	21205 &	0.7 &	17.0	& 1.05 &	0.64 \\ \hline
2.25 & 57877 &		1.125 &	21205 &	0.7 &	14.2	& 2.13 &	0.32	 \\
2.13	& 50281 &		1.125 &	21205 &	0.7 &	14.1	& 2.00 &	0.33	 \\
2.00	& 52489 &		1.125 &	21205 &	0.7 &	14.0	& 1.86 &	0.36	 \\
1.86	& 45037 &		1.125 &	21205 &	0.7 &	13.8	& 1.73 &	0.39	 \\
1.73	& 42721 &		1.125 &	21205 &	0.7 &	13.6	& 1.63 &	0.41 \\
1.63	& 39469 &		1.125 &	21205 &	0.7 &	13.4 & 1.51& 	0.45 \\
1.51	& 40705 &		1.125 &	21205 &	0.7 &	13.2	& 1.38 &	0.49 \\
1.38	& 32929 &		1.125 &	21205 &	0.7 &	13.0	& 1.27 &	0.53 \\
1.27	& 32533 &		1.125 &	21205 &	0.7 &	12.8	& 1.16 &	0.58 \\
1.16	& 28069 &		1.125 &	21205 &	0.7 &	12.7	& 1.06 &	0.64 \\ \hline
2.25 & 57877 &		1.125 &	21205 &	0.5 &	14.2 & 1.85 &	0.37	\\
1.85	& 42337 &		1.125 &	21205 &	0.5 &	13.7	& 1.47 &	0.46	\\
1.47	& 37789 &		1.125 &	21205 &	0.5 &	13.1 & 1.17 &	0.57	\\
1.17	& 33157 &		1.125 &	21205 &	0.5 &	12.5	& 0.97 &	0.67 \\ \hline
2.25	& 57877 &		1.125 &	21205 &	0.5 &	19.0	 & 1.51 & 0.45 \\	
1.51 & 40705 &		1.125 &	21205 &	0.5 & 17.6  &	 1.04 & 0.63 \\ \hline
    \caption{The initial conditions (initial proto-Mercury's and impactor's masses, $M_{p\mercury}$, $M_{imp}$ and resolutions $N_{p\mercury}$, $N_{imp}$, the impact parameter $b$, the impact velocity $v_{imp}$) and outcomes (mass of the largest fragment $M_{lr}$ and its iron mass fraction $Z_{Fe}$) for Case-3 collisions.}
    \label{TableCase3}
   
   \end{longtable}
%

\end{document}